\documentclass[fleqn,usenatbib]{mnras}

\usepackage{newtxtext,newtxmath}

\usepackage[T1]{fontenc}
\usepackage{booktabs}

\usepackage{amssymb}
\usepackage{tabularx, siunitx}
\usepackage[flushleft]{threeparttable}
\usepackage[normalem]{ulem}
\usepackage{makecell}

\DeclareRobustCommand{\VAN}[3]{#2}
\let\VANthebibliography\thebibliography
\def\thebibliography{\DeclareRobustCommand{\VAN}[3]{##3}\VANthebibliography}

\usepackage{graphicx}	
\usepackage{amsmath}	
\usepackage{multirow}   
\usepackage{newtxmath}

\title[Diagnostics of Dyson Sphere Candidate Stars]{Archival Diagnostics for Potential Background Contaminants of \emph{Project Hephaistos} Dyson Sphere Candidates}

\author[T. Ren et al.]{
T. Ren,$^{1}$\thanks{E-mail: tongtian.ren@manchester.ac.uk (TR)}
M. A. Garrett,$^{1,2,7}$, E. Zackrisson,$^{3}$, A. J. Korn,$^{3}$ A. P. V. Siemion,$^{1,4,5,6,7}$ J. T. Wright,$^{8,9,10}$ 
\newauthor and A. Brandeker$^{11}$
\\
$^{1}$ Jodrell Bank Centre for Astrophysics, Department of Physics and Astronomy, University of Manchester, Oxford Road, Manchester M13 9PL, UK\\
$^{2}$ Leiden Observatory, Leiden University, PO Box 9513, NL-2300 RA Leiden, the Netherlands\\
$^{3}$ Observational Astrophysics, Department of Physics and Astronomy, Uppsala University, Box 516, SE-751 20 Uppsala, Sweden\\
$^{4}$ Breakthrough Listen, Astrophysics sub-department, Department of Physics, University of Oxford, Denys Wilkinson Building, Keble Road, Oxford OX1 3RH, UK\\
$^{5}$ Berkeley SETI Research Center, University of California, Berkeley, CA 94720, USA\\
$^{6}$ SETI Institute, 339 Bernardo Avenue, Suite 200, Mountain View, CA 94043, USA\\
$^{7}$ University of Malta, Institute of Space Sciences and Astronomy, Msida, MSD2080, Malta\\
$^{8}$ Department of Astronomy and Astrophysics, The Pennsylvania State University, University Park, PA 16802, USA\\
$^{9}$ Penn State Extraterrestrial Intelligence Center, The Pennsylvania State University, University Park, PA 16802, USA\\
$^{10}$ Center for Exoplanets and Habitable Worlds, The Pennsylvania State University, University Park, PA 16802, USA\\
$^{11}$ Department of Astronomy, Stockholm University, AlbaNova University Center, 10691 Stockholm, Sweden}

\date{Accepted by the Monthly Notices of the Royal Astronomical Society (MNRAS).}

\pubyear{\the\year{}}

\begin{document}

\label{firstpage}
\pagerange{\pageref{firstpage}--\pageref{lastpage}}
\maketitle

\begin{abstract}
We report on the diagnostic investigation of nine \emph{Project Hephaistos} Dyson Sphere candidate M-dwarfs based on archival data. By comparing the \textit{Gaia} DR3 positions with epoch 2016.0, propagated to the ALLWISE epoch, with the mid-infrared centroids measured from the ALLWISE images, together with deep archival optical/near-infrared imaging, we identified significant background contamination in candidates B and C---candidate B coincident with a radio counterpart showing a power-law-like spectrum with a radio spectral index $\alpha = 0.63 \pm 0.11$, and candidate C with a near-infrared companion at a $3.50\arcsec$ offset. Candidate A provides suggestive evidence through a radio counterpart with $\alpha = 0.40 \pm 0.35$, while candidates H and J show some marginal evidence. These systems exhibit either significant astrometric offsets or visible interlopers, indicating that the mid-infrared excess likely arises from line-of-sight contamination by hot, dust-obscured galaxies. However, candidates D, E, F, and I still lack obvious signs of contamination. Dedicated observations are therefore essential to characterise these potential interlopers, in order to eliminate false positives and ensure that techno-signature searches focus on the most robust Dyson Sphere candidates.


\end{abstract}




\section{Introduction}

In mid-2024, \emph{Project Hephaistos}\footnote{\url{https://www.astro.uu.se/~ez/hephaistos/hephaistos.html}}
reported seven Dyson sphere candidates. Dyson spheres are hypothetical megastructures envisioned to harvest stellar energy, with their most distinctive observable signature being excess infrared emission produced by the re-radiation of absorbed starlight as thermal waste heat \citep{1960Sci...131.1667D, 2020SerAJ.200....1W}. The project identified these candidates by selecting stars with significant mid-infrared (MIR) excess from a sample of approximately 5 million stars detected by \emph{Gaia}, 2MASS, and \emph{WISE}, all within 300 parsecs of us \citep{2024MNRAS.531..695S}. All candidates are M-dwarf stars exhibiting MIR excess in the \emph{WISE} $W3$ and $W4$ bands, with spectral energy distributions (SEDs) consistent with blackbody temperatures in the range of 100--300~K, as expected from partial Dyson Sphere models \citep{2022MNRAS.512.2988S}.

Just as radio SETI surveys targeting nearby stars frequently encompass background extragalactic objects within the telescope’s field of view \citep{2023MNRAS.519.4581G}, the same principle applies analogously to infrared SETI searches. The cross-matching of seven candidates with archival radio surveys revealed offset radio counterparts for candidates A, B, and G \citep*{2024RNAAS...8..145R}. High-resolution e-MERLIN (1.5 GHz) and EVN (5 GHz) observations \citep*{2025MNRAS.538L..56R} confirmed that J2335$-$0004, the radio source associated with candidate G, is a high brightness temperature ($\rm > 1\times 10^8~K$) background active galactic nucleus (AGN) located more than 5 arcseconds from the propagated epoch 2024.4 \emph{Gaia} position of the star. We did not detect radio emission exceeding $3\sigma$ at the stellar coordinates at 1.5 or 5~GHz, and thus found no evidence of broadband radio emission at these frequencies that could represent unintentional radio leakage from the candidate system.

Furthermore, the positional offset analysis of the ALLWISE images revealed that while the emission in the \emph{WISE} $W1$ and $W2$ bands is associated with the M-dwarf, the emission in the $W3$ and $W4$ bands is dominated by the offset AGN. The spectral energy distribution (SED) of the AGN suggested that it was most likely a hot dust-obscured galaxy (hot DOG; \citealt{2015ApJ...804...27A,2015ApJ...805...90T}), which is bright in the $W3$ and $W4$ bands but faint and obscured in the $W1$ and $W2$ bands. 

Hot DOGs are a rare class of luminous galaxies powered by heavily obscured supermassive black holes, representing a distinct phase in galaxy evolution. Due to severe dust obscuration, they appear faint in the optical and near-infrared (NIR) but exhibit strong emission in the mid-infrared (MIR) as heated dust re-radiates energy. Consequently, hot DOGs are faint in the \emph{WISE} $W1$ and $W2$ bands but prominent in the $W3$ and $W4$ bands. When coincident with nearby M-dwarfs along the line of sight, the coarse 6--12~arcsec resolution of \emph{WISE} results in blended photometry that can mimic the SED of a cold Dyson Sphere. Notably, \citet{2024arXiv240911447B} estimated the hot DOG occurrence rate in the \emph{WISE} catalogue, concluding that their surface density is sufficient to explain all seven reported Dyson Sphere candidates as chance alignments.

\defcitealias{2024MNRAS.531..695S}{S24}
\defcitealias{2026arXiv260725701K}{K26}

For the remaining six Dyson Sphere candidates of \citet{2024MNRAS.531..695S}, two have radio counterparts \citep*{2024RNAAS...8..145R} that are also likely to be AGNs (candidates A and B). The nature of the other four (candidates C, D, E, and F) is unclear. Subsequently, further identification of the optical spectra of the \emph{Project Hephaistos} stars led to the ancillary discovery of three additional Dyson Sphere candidates (H, I and J), which show significant MIR excesses in the $W3$ and $W4$ bands (\citet{alma991018707829107596} and \citet{2026arXiv260725701K}).

Beyond searching for typical AGN signatures in potential background contaminants (\textit{e.g.} high brightness temperature radio and X-ray emission), a complementary method involves identifying significant astrometric offsets across wavebands. By comparing \emph{WISE} centroid positions with high-precision \emph{Gaia} astrometry (accurate to microarcsecond scales), we can detect spatial offsets that serve as clear indicators of background contamination. This effect was explicitly observed in the analysis of candidate G. Crucially, however, this approach relies on the background contaminant being spatially distinct from the foreground star; if the alignment is sufficiently precise, such offsets will remain unmeasurable. While \citet{2024MNRAS.531..695S} used $W1$- and $W3$-band offsets to evaluate discrepancies and employed the Improved Reprocessing of the \emph{IRAS} Survey (IRIS; \citealt{2005ApJS..157..302M}) maps at 100~$\mu$m to mitigate potential contamination, residual contamination remains in some cases, as demonstrated by the radio observations of candidate G \citep*{2025MNRAS.538L..56R}. Owing to the limited spatial resolution of \emph{WISE} and \emph{IRAS}, directly measuring the offset between MIR centroids across the four \emph{WISE} bands and \emph{Gaia}-measured stellar positions provides a more effective and precise diagnostic for identifying extragalactic contamination in apparent infrared-excess sources, yielding the best constraint attainable from existing data before high-resolution imaging observations.

Supplementing the primary data from \emph{Gaia} \citep{2016A&A...595A...1G, 2023A&A...674A...1G}, 2MASS \citep{2006AJ....131.1163S}, and \emph{WISE} \citep{2010AJ....140.1868W}, deeper archival imaging surveys cover several of the nine targets.
These include optical surveys such as the Dark Energy Survey (DES; \citealp{2016MNRAS.460.1270D}; DES DR1: \citealp{2018ApJS..239...18A}; DES DR2: \citealp{2021ApJS..255...20A}), the Panoramic Survey Telescope and Rapid Response System (Pan-STARRS; \citealp{2016arXiv161205560C}), SkyMapper \citep{2024PASA...41...61O}, and the DESI Legacy Imaging Surveys \citep{2019AJ....157..168D}, as well as NIR surveys such as the VISTA Science Archive \citep{2012A&A...548A.119C} and the WFCAM Science Archive \citep{2008MNRAS.384..637H}. The greater depth of these images provides a valuable independent diagnostic, offering clues to potential faint background extragalactic objects contaminating the MIR flux of the Dyson Sphere candidates. The culprits responsible for the excess, though absent in shallower surveys like SDSS and 2MASS, may be detectable in these high-sensitivity archival images.

This paper presents a comprehensive assessment of the Dyson Sphere candidates through two complementary approaches: an astrometric diagnosis based on positional offsets and a detailed inspection of deep photometric surveys. The paper is structured as follows: in Section \ref{sec:section_2}, we briefly describe the data and methods used in this paper. We present the results of our astrometric diagnostics in Section \ref{sec:section_3}, followed by the archival image examination results in Section \ref{sec:section_4}. Finally, we discuss our findings in Section \ref{sec:section_5} and summarise our conclusions in Section \ref{sec:section_6}.

\section{Data and Methods}
\label{sec:section_2}

\subsection{\emph{Gaia} Astrometry, Radio Counterparts, and \emph{WISE} Centroid Measurement}

\begin{table*}
\centering
\caption{Basic \emph{Gaia} DR3 parameters of the Dyson Sphere candidate stars 
from \citet{2024MNRAS.531..695S}, \citet{alma991018707829107596}, and \citet{2026arXiv260725701K}.}
\label{tab:candidates}
\fontsize{6.5pt}{9pt}\selectfont
\begin{tabular}{cccccccccccc}
\hline
\makecell{Candidate} & \makecell{\emph{Gaia} DR3} & \makecell{RA, DEC\\(J2016.0, deg)} & \makecell{$G$ (Vega)\\{(mag)}} & \makecell{$T_{\mathrm{eff}}$\\{(K)}} & \makecell{pmRA\\{(mas\,yr$^{-1}$)}} & \makecell{pmDEC\\{(mas\,yr$^{-1}$)}} & \makecell{PM\\{(mas\,yr$^{-1}$)}} & \makecell{Plx\\{(mas)}} & \makecell{$W1$/$W2$\\epoch} & \makecell{$W3$/$W4$\\epoch} & \makecell{NIR\\survey} \\
\hline
A (J1245-2652) & 3496509309189181184 & 191.30392, $-$26.86758 & 15.996 & 3823 & -5.717  & -88.760 & 88.944 & 6.93 & 2010.52 & 2010.51 & VHS ($YJK_{\rm s}$) \\
B (J0356-4031) & 4843191593270342656 & 59.01583, $-$40.53001  & 17.713 & 3431 & 7.051   & -31.500 & 32.279 & 4.68 & 2010.59 & 2010.58 & VHS ($JK_{\rm s}$) \\
C (J0456-7410) & 4649396037451459712 & 74.01205, $-$74.17051  & 18.393 & 3239 & 39.525  & -19.288 & 43.980 & 4.54 & 2010.78 & 2010.29 & VMC ($YJK_{\rm s}$) \\
D (J2327+0506) & 2660349163149053824 & 351.96373, $+$5.10726  & 17.662 & 3473 & -30.674 & -21.611 & 37.522 & 4.70 & 2010.46 & 2010.46 & UKIDSS ($YJHK$) \\
E (J0402-1054) & 3190232820489766656 & 60.53249, $-$10.91131  & 17.008 & 3557 & 33.957  & 30.112  & 45.385 & 3.62 & 2010.62 & 2010.11 & VHS ($YJK_{\rm s}$) \\
F (J0513-2511) & 2956570141274256512 & 78.44374, $-$25.18643  & 16.330 & 3674 & 11.575  & -20.615 & 23.643 & 3.73 & 2010.16 & 2010.16 & VHS ($JK_{\rm s}$) \\
G (J2335-0004) & 2644370304260053376 & 353.88537, $-$0.07339  & 16.482 & 3481 & -47.485 & -49.646 & 68.699 & 3.97 & 2010.46 & 2010.46 & VHS ($JHK_{\rm s}$) \\
H (J2336-0920) & 2437221214075471744 & 354.01111, $-$9.33344  & 17.312 & 3418 & -23.535 & -15.753 & 28.321 & 3.68 & 2010.45 & 2010.45 & VHS ($YJK_{\rm s}$) \\
I (J0939+0700) & 3854090071297359616 & 144.97633, $+$7.00774  & 17.402 & 3327 & -4.744  & -14.760 & 15.504 & 5.89 & 2010.36 & 2010.35 & UKIDSS ($YJHK$) \\
J (J0832+1442) & ~~651765552072217216 & 128.06908, $+$14.70515 & 16.133 & 3677 & -5.615  & -30.893 & 31.399 & 3.42 & 2010.82 & 2010.30 & UHS ($JHK$) \\
\hline
\end{tabular}
\begin{flushleft}
    \small
    \textit{Note.} --- All the candidates have complete \emph{Gaia} astrometry and \emph{WISE} photometry data. $T_{\mathrm{eff}}$ (Effective Temperature) is not available in \emph{Gaia} DR3 for candidate A; we adopted its $T_{\mathrm{eff}}$ from \emph{Gaia} DR2 instead. All the right ascension and declination positions are given in decimal degrees (ICRS) for the \emph{Gaia} DR3 reference epoch J2016.0; $\mathrm{pmRA}$, $\mathrm{pmDEC}$, PM, and $\mathrm{Plx}$ denote the proper-motion components in right ascension and declination, the total proper motion, and the trigonometric parallax, respectively, as reported in \emph{Gaia} DR3. All $G$-band magnitudes are reported on the Vega system. All candidates are covered by Legacy imaging in the $g$, $r$, $i$, and $z$ bands, with the exception of candidate E, for which only the $g$, $r$, and $z$ bands are available; the final column therefore lists only the NIR survey used for each candidate. The radio band information for A and B is described in Table~\ref{tab:AB_radio}.
\end{flushleft}
\end{table*}

We collected public archival data for nine M-dwarf candidates: six from \citet{2024MNRAS.531..695S} (\citetalias{2024MNRAS.531..695S} hereafter) and three from \citet{alma991018707829107596} and \citet{2026arXiv260725701K} (\citetalias{2026arXiv260725701K} hereafter), all of which have \emph{Gaia} DR3 source identifiers (see Table~\ref{tab:candidates}). We retrieved their high-precision astrometry, including coordinates and proper motions, for all nine stars from the \emph{Gaia} DR3 database. Using these positions, we searched for counterparts at complementary wavelengths. Radio counterparts for candidates~A and B were found in the predecessor identification \citep*{2024RNAAS...8..145R} via the VizieR access tool \citep{2000A&AS..143...23O}, while candidates C--F and X-ray counterparts for all candidates remained undetected. In this work, we updated the VizieR query to account for new radio survey data and cross-matched all positions with the High Energy Astrophysics Science Archive Research Center (HEASARC) database\footnote{\url{https://heasarc.gsfc.nasa.gov/}}, which similarly yielded no additional detections.

To prepare the data for MIR centroid analysis in Section~\ref{sec:section_3}, we retrieved $1\arcmin \times 1\arcmin$ image cutouts for each star in all four \emph{WISE} bands from the ALLWISE atlas \citep{2020ipac.data.I153W} via the NASA/IPAC Infrared Science Archive (IRSA)\footnote{\url{https://irsa.ipac.caltech.edu/irsaviewer/}}. To refine the centroid analysis, we manually measured the signal-to-noise ratios (SNRs) for the four ALLWISE band images of each source, rather than adopting the pipeline-generated SNR values in the ALLWISE catalogue \citep{2014yCat.2328....0C}. We used the \texttt{sep.extract} method from the \textsc{SEP} Python library \citep{barbary2016sep}, a re-implementation of \textsc{SExtractor} \citep{1996A&AS..117..393B}, to detect the centroids of the ALLWISE maps. The $W1$ and $W2$ images, which have significantly higher SNRs, could generally be fitted successfully with \textsc{SEP}. We evaluated several alternative methods for detecting WISE spatial centroids and determined that \textsc{SEP} generally yields the best performance. However, for lower-resolution $W3$ and $W4$ images for which \textsc{SEP} failed to provide a successful fit, we adopted the \texttt{scipy.ndimage.center\_of\_mass} function.

Neither \texttt{sep.extract} nor \texttt{scipy.ndimage.center\_of\_mass} provides direct centroid position uncertainties. We estimated these using the standard photon-limited centroiding precision formula:

\begin{equation}
\sigma_{\rm position} \approx \frac{\text{FWHM}}{k \times \text{SNR}}
\end{equation}

where ${k = 2\sqrt{2\ln 2} \approx 2.355}$ \citep{2006hca..book.....H} converts between Full Width at Half Maximum (FWHM) and Gaussian ${\sigma}$. We adopted the \emph{WISE} point spread function (PSF) FWHM of 6.1, 6.4, 6.5, and 12.0 arcseconds for the $W1$--$W4$ bands, respectively \citep{2010AJ....140.1868W}. The SNR at each centroid was manually measured via \texttt{SEP} aperture photometry. We derived all centroiding uncertainties (${\sigma_{\rm{position}}}$) using this formula.

\subsection{Epoch Propagation and Proper Motion Corrections}

The extracted \emph{WISE} centroids and their associated positional uncertainties reflect the state of the MIR emission at the specific epoch of the observations. As summarised in Table~\ref{tab:candidates}, the ALLWISE epochs for our targets span J2010.0 to J2011.0. Even within the same survey, the $W3$- and $W4$-band data were often collected slightly earlier than the $W1$- and $W2$-band data. Because our multi-wavelength dataset compiles images from multiple independent surveys, these epoch differences can be substantial. Consequently, we rigorously accounted for the proper motions of the target stars when examining and cross-matching the spatial centroids in Section~\ref{sec:section_3}.

In this context, the \emph{Gaia} parallaxes in Table~\ref{tab:candidates} span 3.42--6.93~$\mathrm{mas}$ and are negligible on the angular scales considered here. In contrast, the total proper motions range from 15.5 to 88.9~$\mathrm{mas\,yr^{-1}}$, with a median of 34.9~$\mathrm{mas\,yr^{-1}}$, corresponding to a displacement of approximately 0.35$\arcsec$ over 10 years. For instance, candidate~A has a significant proper motion of $88.944~{\rm mas\,yr^{-1}}$. While its RACS-high radio image was obtained at J2022.518, its MIR images were acquired by \emph{WISE} near J2010.5. This $\sim$12\,yr baseline introduces a purely kinematic positional displacement of $1.07\arcsec$.

The \emph{Gaia} DR3 astrometric solution is based on observations collected between 2014 and 2017, with the catalogue positions and proper motions referenced to epoch 2016.0. We therefore propagate the \emph{Gaia} coordinates to the observing epoch of each individual image, preventing stellar proper motion from introducing spurious positional offsets. By applying this temporal alignment to every dataset, the uncertainty in each derived offset includes the centroiding uncertainty ($\sigma_{\mathrm{position}}$), together with the propagated positional and proper-motion uncertainties of the \emph{Gaia} stellar source and, where applicable, the positional uncertainty of the corresponding radio counterpart.

\subsection{Deep Optical and Near-Infrared Imaging Surveys}

To examine the candidates using deep photometric and imaging data in Section~\ref{sec:section_4}, we queried several wide-field optical and NIR surveys, including the DES DR2, Pan-STARRS DR2, DESI Legacy Imaging Surveys (DR10), the UKIRT Infrared Deep Sky Survey (UKIDSS; \citealp{2007MNRAS.379.1599L}) and the UKIRT Hemisphere Survey (UHS; \citealp{2018MNRAS.473.5113D}) from the WFCAM Science Archive \footnote{\url{http://wsa.roe.ac.uk/}}, and two VISTA programmes from the VISTA Science Archive \footnote{\url{http://vsa.roe.ac.uk/}}: the Hemisphere Survey (VHS; DR5, \citet{2013Msngr.154...35M}) and the Magellanic Clouds Survey (VMC; DR6, \citet{2011A&A...527A.116C}). Table~\ref{tab:obsdata} summarises the central/effective wavelengths (or frequencies) and nominal angular resolutions of the optical and infrared datasets used in this work, together with the RACS radio surveys \citep{2021PASA...38...58H,2024PASA...41....3D,2025PASA...42...38D} covering candidates A and B. We find that all candidates are covered by the Legacy DR10 Survey, which provides the ideal optical imaging and serves as the primary dataset for this investigation. Additionally, six candidate stars have VISTA imaging: A, B, E, F, and H from VHS, and C from VMC. The remaining candidates lack VISTA coverage and are instead covered by UKIRT surveys: D and I by UKIDSS, and J by UHS. These Legacy optical images, combined with the NIR data from VISTA or UKIRT surveys, provide deep, multi-band imaging across all candidate fields. This enables a visual diagnosis entirely independent of our mid-infrared centroid analysis.

\begin{table*}
\centering
\caption{Summary of the observational datasets used in this work. For each band we give the effective wavelength or central frequency and the nominal angular resolution, quoted as the full width at half-maximum (FWHM) of the synthesised beam for the radio surveys and of the point-spread function (PSF) for the optical and infrared imaging data. The final column gives the reference for each survey.}
\label{tab:obsdata}
\begin{tabular}{lcccl}
\hline
\makecell{Survey\\(instrument)} &
\makecell{Band} &
\makecell{Effective Wavelength /\\Central Frequency\\
($\lambda_{\rm eff}/\nu_{\rm c}$)} &
\makecell{FWHM\\($\arcsec$)} &
\makecell{Reference} \\
\hline
\multirow{3}{*}{ASKAP RACS} & RACS-low  & 887.5\,MHz  & $\sim$25 & \citet{2021PASA...38...58H} \\
                            & RACS-mid  & 1367.5\,MHz & $\sim$10 & \citet{2024PASA...41....3D} \\
                            & RACS-high & 1655.5\,MHz & $\sim$10 & \citet{2025PASA...42...38D} \\
\hline
\emph{Gaia} (Astrometric Field) & $G$ & 0.64\,$\mu$m & $\sim$0.10 & \mbox{\citet{2016A&A...595A...3F,2018A&A...617A.138W}} \\
\hline
\multirow{4}{*}{Legacy Survey} & $g$ & 0.48\,$\mu$m & 1.3 & \multirow{4}{*}{\citet{2019AJ....157..168D}} \\
                               & $r$ & 0.64\,$\mu$m & 1.2 & \\
                               & $i$ & 0.78\,$\mu$m & $\sim$1.1--1.2 & \\
                               & $z$ & 0.92\,$\mu$m & 1.1 & \\
\hline
\multirow{4}{*}{VISTA (VMC, VHS)} & $Y$   & 1.02\,$\mu$m & \multirow{4}{*}{$\sim$0.8--1.0} & \multirow{4}{*}{\citet{2011A&A...527A.116C,2013Msngr.154...35M}} \\
                                  & $J$   & 1.25\,$\mu$m &  & \\
                                  & $H$   & 1.65\,$\mu$m &  & \\
                                  & $K_{\rm s}$ & 2.15\,$\mu$m &  & \\
\hline
\multirow{4}{*}{UKIRT/WFCAM (UKIDSS, UHS)} & $Y$ & 1.03\,$\mu$m & \multirow{4}{*}{$\sim$0.75--1.0} & \multirow{4}{*}{\citet{2007MNRAS.379.1599L,2018MNRAS.473.5113D}} \\
                                           & $J$ & 1.25\,$\mu$m &  & \\
                                           & $H$ & 1.63\,$\mu$m &  & \\
                                           & $K$ & 2.20\,$\mu$m &  & \\
\hline
\multirow{4}{*}{AllWISE} & $W1$ & 3.4\,$\mu$m & 6.1  & \multirow{4}{*}{\citet{2010AJ....140.1868W}} \\
                         & $W2$ & 4.6\,$\mu$m & 6.4  & \\
                         & $W3$ & 12\,$\mu$m  & 6.5  & \\
                         & $W4$ & 22\,$\mu$m  & 12.0 & \\
\hline
\end{tabular}
\end{table*}

To provide identifiers for newly discovered or faint objects in each candidate’s images, we queried nearby NIR sources from the VISTA and UKIRT catalogues and optical sources from the Legacy DR10 \textsc{Tractor} catalogue. We also examined other major optical and NIR surveys, including DES, Pan-STARRS, SDSS, 2MASS, and SkyMapper. The DES images are already included in the DESI Legacy DR10 Imaging Surveys and thus are effectively represented in our Legacy analysis. The other surveys---Pan-STARRS, SDSS, 2MASS, and SkyMapper---have imaging quality and depth that are generally inferior to those of the Legacy, VISTA, and UKIRT datasets in the regions of interest. Our follow-up examination in Section \ref{sec:section_4} based on them did not yield additional detections of background or blended sources.


\section{Astrometric Diagnostics}
\label{sec:section_3}

In this section, we analyse the positional offset between the MIR emission and the actual \emph{Gaia} position for each candidate star. As demonstrated by \citet*{2025MNRAS.538L..56R}, which analysed the case of candidate G, a significant offset between the spatial distributions of the $W3$ and $W4$ emission and the stellar coordinates is an indicator of background contamination. For clarity, we divide our nine candidates into three subsamples based on their properties and origin:

\begin{itemize}
    \item Candidates A and B, which are found to have radio counterparts. We additionally compare the radio counterpart positions with the WISE centroids and the \emph{Gaia} positions of the candidate stars.
    \item Candidates C, D, E, and F, the other four candidates drawn from \citetalias{2024MNRAS.531..695S} without radio counterparts.
    \item Candidates H, I, and J, the three newly identified targets from \citetalias{2026arXiv260725701K}, also without radio counterparts.
\end{itemize}

The diagnostic process and results for each group are detailed in Subsections \ref{sec:section_3_1}, \ref{sec:section_3_2}, and \ref{sec:section_3_3}. In Section \ref{sec:section_3_4}, we compare the candidate centroiding results in this work with the offsets reported in \citetalias{2024MNRAS.531..695S}. We then evaluate our astrometric approach in Subsection~\ref{sec:section_3_5} by analysing control samples with $G$-mag , $W1$- and $W2$-band SNRs similar to those of the candidate stars.

\subsection{Radio, WISE and \emph{Gaia} astrometric diagnostics for candidates A and B}
\label{sec:section_3_1}

\begin{figure*}
    \centering
    \includegraphics[width=1.0\linewidth]{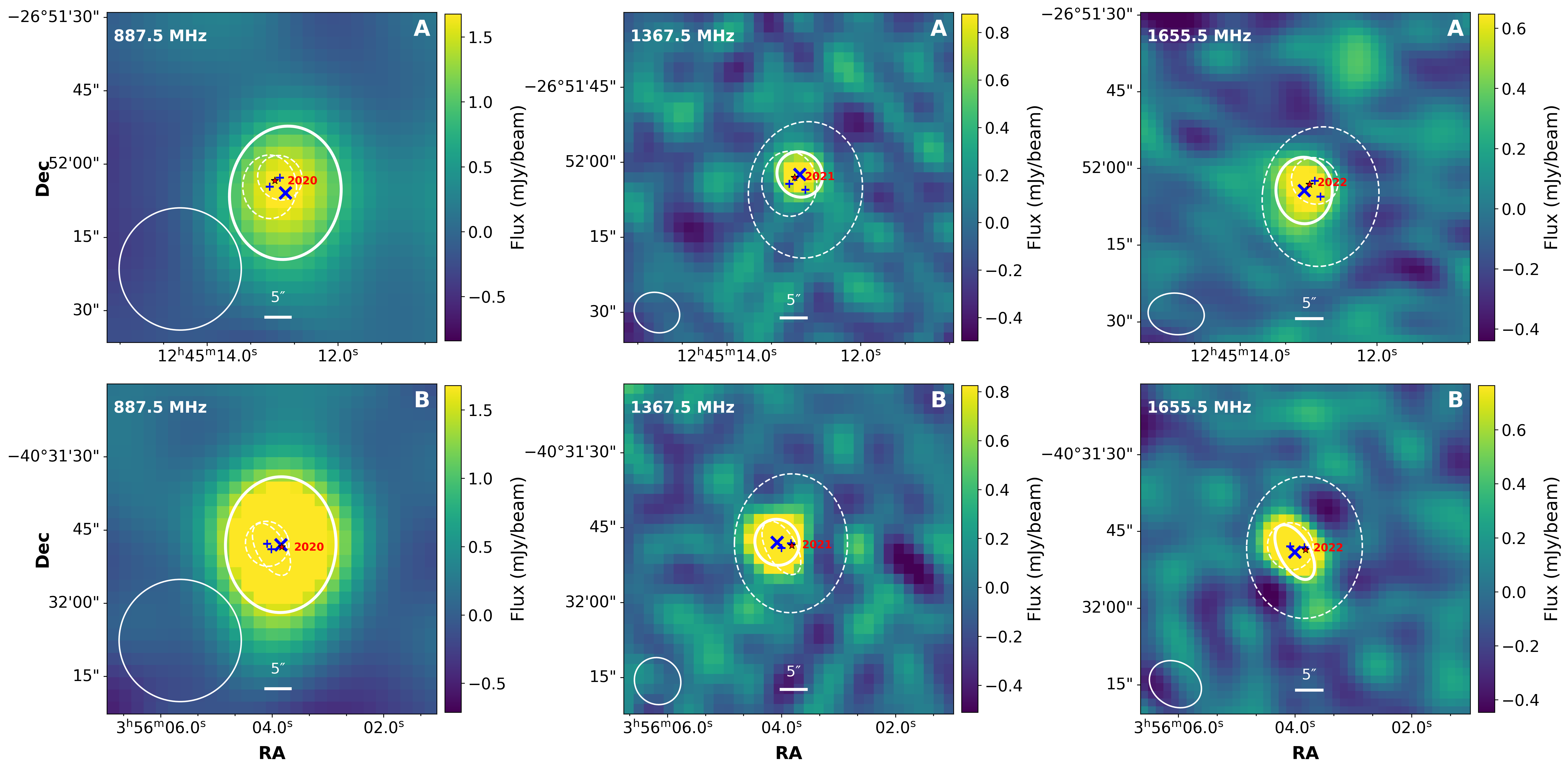}
    \caption{The Rapid ASKAP Continuum Survey (RACS) radio images of radio sources associated with candidates A (J1245-2652, top row) and B (J0356-4031, bottom row) are shown at three frequencies: 887.5 MHz (left), 1367.5 MHz (centre), and 1655.5 MHz (right), with each panel covering a field of view of 65 $^{\arcsec} \times$ 65 $^{\arcsec}$. White ellipses show the PyBDSF source fits from the RACS catalogues, with thick solid lines indicating the fits at the frequency of the image, while thin dashed lines show fits from the other radio frequencies for comparison. Large blue crosses ($\times$) mark radio source positions for the corresponding frequency, and small blue plus signs (+) indicate positions at other frequencies. The red stars represent the \emph{Gaia} positions propagated to the corresponding RACS observation epochs using the measured proper motions.}
    \label{fig:AB_radio}
\end{figure*}

Candidates A (\emph{Gaia} DR3 3496509309189181184) and B (\emph{Gaia} DR3 4843191593270342656) are two of the three radio-associated sources (A, B, and G) reported by \citet*{2024RNAAS...8..145R}; the radio source associated with G has been confirmed as an AGN \citep*{2025MNRAS.538L..56R}. Candidates A and B are located in the southern sky, with associated radio sources detected in RACS-DR1 \citep{2021PASA...38...58H} at 887.5~MHz with an angular resolution of 25${\arcsec}$. More recent RACS-mid \citep{2024PASA...41....3D} and RACS-high \citep{2025PASA...42...38D} observations at 1367.5~MHz and 1655.5~MHz provided significantly improved angular resolution compared to the original 25$^{\arcsec}$ beam observations, enabling more precise comparison between radio source positions and stellar coordinates from \emph{Gaia}. Figure~\ref{fig:AB_radio} shows the positions of the radio sources associated with candidates A and B at different radio frequencies, and Table~\ref{tab:AB_radio} lists their properties.

\begin{figure}
    \centering
    \includegraphics[width=1\linewidth]{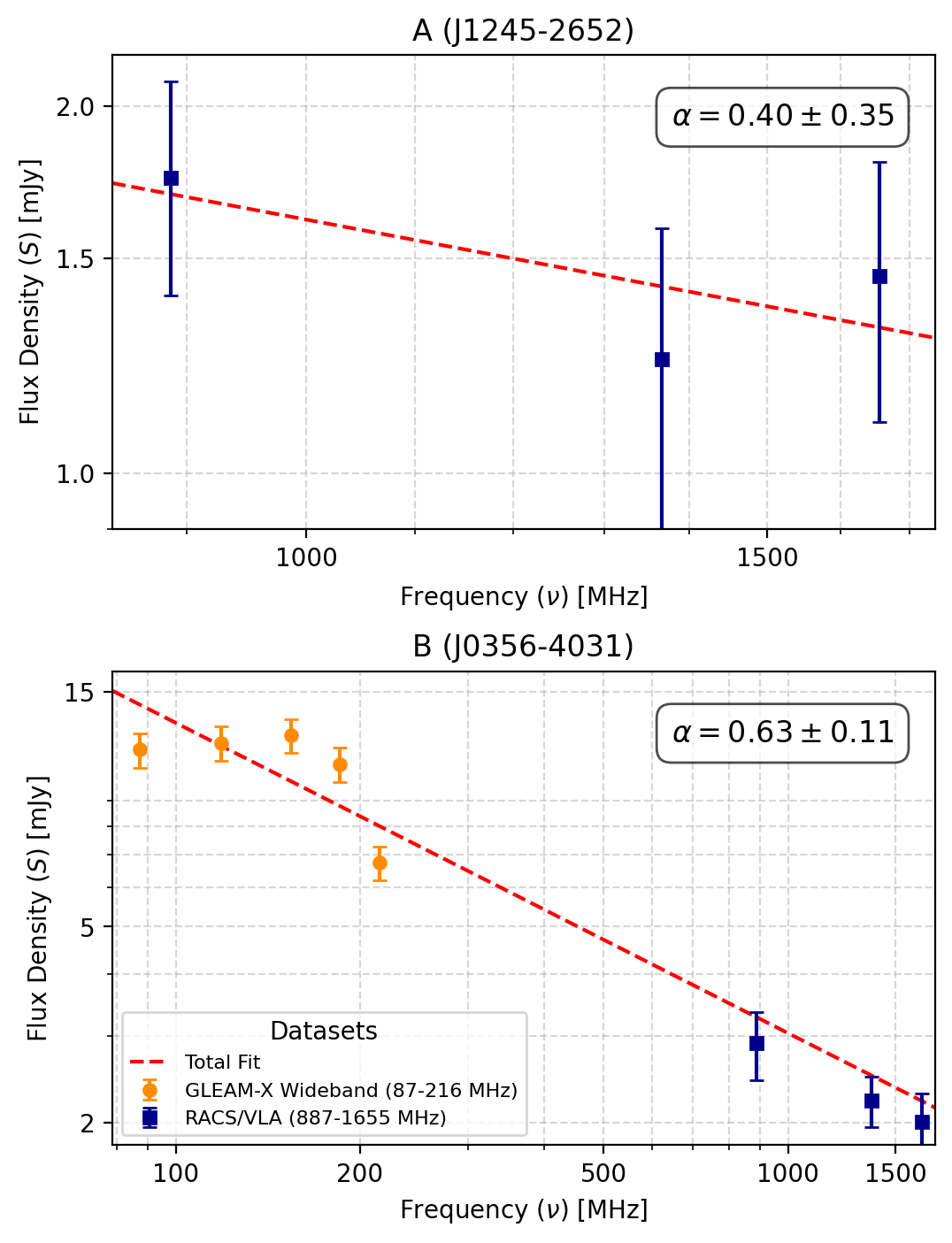}
    \caption{Radio spectral energy distributions and spectral index fits for candidates A (J1245-2652, top panel) and B (J0356-4031, bottom panel). Dark blue squares represent RACS-low, RACS-mid, and RACS-high observations at 887.5, 1367.5, and 1655.5~MHz. Orange circles in the bottom panel show GLEAM-X wideband integrated fluxes across five frequency channels (87.5, 118.5, 154.5, 185.5, and 215.5~MHz). The red dashed lines show power-law fits to the data, with spectral indices ($\alpha$) derived via weighted least-squares fitting in log-space.}
    \label{fig:alpha}
\end{figure}

Another supplementary dataset is the second data release for the GaLactic and Extragalactic All-sky Murchison Widefield Array eXtended (GLEAM-X) \citep{2024PASA...41...54R} which covers a frequency range of 72--231 MHz with twenty frequency bands. Although the angular resolution of GLEAM-X is at the arcminute level, it still effectively captures the fluxes of candidate B at lower frequencies. Combining GLEAM-X and RACS data extends the coverage from 87.5 to 1655.5~MHz. J0356-4031, the radio source associated with candidate B, has a high flux density exceeding 10~mJy at frequencies below 300~MHz and shows a steep radio spectral index (assuming a power-law radio spectrum $S_{\nu} \propto \nu^{-\alpha}$, where $S_{\nu}$ is the flux density, $\nu$ is the frequency, and $\alpha$ is the spectral index) of $\alpha^{1655.5}_{87.5} =0.63\pm 0.11$ (Figure~\ref{fig:alpha}). The radio source J1245-2652 associated with candidate A, detected only in RACS, has a less well-constrained radio spectral index of 
$\alpha^{1655.5}_{887.5} = 0.40\pm 0.35$.

\begin{table*}
\centering
\scriptsize
\caption{Properties of the radio sources associated with candidates A and B.}
\label{tab:AB_radio}
\begin{tabular}{cccccccccc}
\hline
\makecell{Source} &
\makecell{Survey\\(Epoch)} &
\makecell{Central Freq.\\(MHz)} &
\makecell{RA (J2000)\\(h m s)} &
\makecell{Dec (J2000)\\($^\circ$ $^\prime$ $^{\prime\prime}$)} &
\makecell{Flux Density\\(mJy)} &
\makecell{Beam size\\
($\arcsec\times\arcsec$ ($^\circ$))} &
\makecell{PyBDSF fit\\
($\arcsec\times\arcsec$ ( $^\circ$))} &
\makecell{Offset to star\\($^{\prime\prime}$)} &
\makecell{Weighted mean\\offset ($^{\prime\prime}$)} \\
\hline
A & RACS-low (J2020.7917) & 887.5 & 12 45 12.783 & $-$26 52 06.20 & $1.75 \pm 0.40$ & 25.0 $\times$ 25.0 (0.0) & 27.4 $\times$ 22.8 (171.8) & $3.26 \pm 2.03$ & \multirow{3}{*}{$1.27 \pm 0.34$} \\
 (J1245-2652) & RACS-mid (J2021.0490) & 1367.5 & 12 45 12.864 & $-$26 52 03.11 & $1.24 \pm 0.31$ & 9.3 $\times$ 7.9 (67.3) & 9.4 $\times$ 8.8 (45.0) & $1.18 \pm 0.36$ & \\
 & RACS-high (J2022.0518) & 1655.5 & 12 45 13.022 & $-$26 52 04.94 & $1.45 \pm 0.39$ & 11.0 $\times$ 8.1 (82.9) & 13.0 $\times$ 11.0 (2.0) & $1.59 \pm 1.24$ & \\
\hline
 & & 87.5 &  &  & 11.43 & 186.9 $\times$ 152.4 ($-$53.2) & 188.2 $\times$ 151.4 ($-$54.0) &  & \multirow{8}{*}{$2.64 \pm 0.36$} \\
 &  & 118.5 &  &  & 11.81 & 123.5 $\times$ 97.6 ($-$53.9) & 123.5 $\times$ 96.0 ($-$54.0) &  & \\
 & GLEAM-X (J2020.0) & 154.5 & 03 56 03.86 & $-$40 31 48.8 & 12.27 & 95.1 $\times$ 74.3 ($-$54.9) & 97.5 $\times$ 72.2 ($-$54.0) & $0.58 \pm 4.07$ & \\
B &  & 185.0 &  &  & 10.74 & 77.5 $\times$ 61.9 ($-$56.0) & 80.5 $\times$ 59.4 ($-$54.0) &  & \\
(J0356-4031) &  & 215.5 &  &  & 6.79 & 65.7 $\times$ 52.7 ($-$55.0) & 69.1 $\times$ 49.7 ($-$54.0) &  & \\
\cline{2-9}
 & RACS-low (J2020.7924) & 887.5 & 03 56 03.831 & $-$40 31 48.19 & $2.90 \pm 0.46$ & 25.0 $\times$ 25.0 (0.0) & 27.8 $\times$ 22.6 (178.3) & $0.53 \pm 1.41$ & \\
 & RACS-mid (J2021.0179) & 1367.5 & 03 56 04.074 & $-$40 31 48.00 & $2.22 \pm 0.29$ & 9.6 $\times$ 9.1 (37.6) & 9.3 $\times$ 8.8 (25.0) & $3.05 \pm 0.36$ & \\
 & RACS-high (J2022.0645) & 1655.5 & 03 56 03.998 & $-$40 31 49.12 & $2.01 \pm 0.35$ & 10.7 $\times$ 8.6 (58.8) & 11.8 $\times$ 6.0 (29.2) & $2.14 \pm 0.50$ & \\
\hline
\end{tabular}
\begin{flushleft}
\textit{Note.} — Beam sizes are the synthesised beam FWHM, given as major axis $\times$ minor axis ($^{\prime\prime}$ $\times$ $^{\prime\prime}$) and position angle (PA, $^\circ$). PyBDSF fit sizes are the fitted source parameters (convolved with beam) from RACS catalogues \citep{2021PASA...38...58H,2024PASA...41....3D,2025PASA...42...38D} and GLEAM-X \citep{2024PASA...41...54R}. The 'Offset to star' is the angular offset between the radio source position and the proper-motion- and epoch-corrected \emph{Gaia} DR3 position of the star. Radio-to-stellar offsets for GLEAM-X are calculated using the propagated \emph{Gaia} position at epoch $\sim$2020.0; radio-to-stellar offsets for RACS are based on their epochs provided by the corresponding source catalogues. The 'Weighted mean offset' column gives the inverse-variance weighted average of the per-band offsets for each candidate.
\end{flushleft}
\end{table*}

\begin{figure*}
    \centering
    \includegraphics[width=1\linewidth]{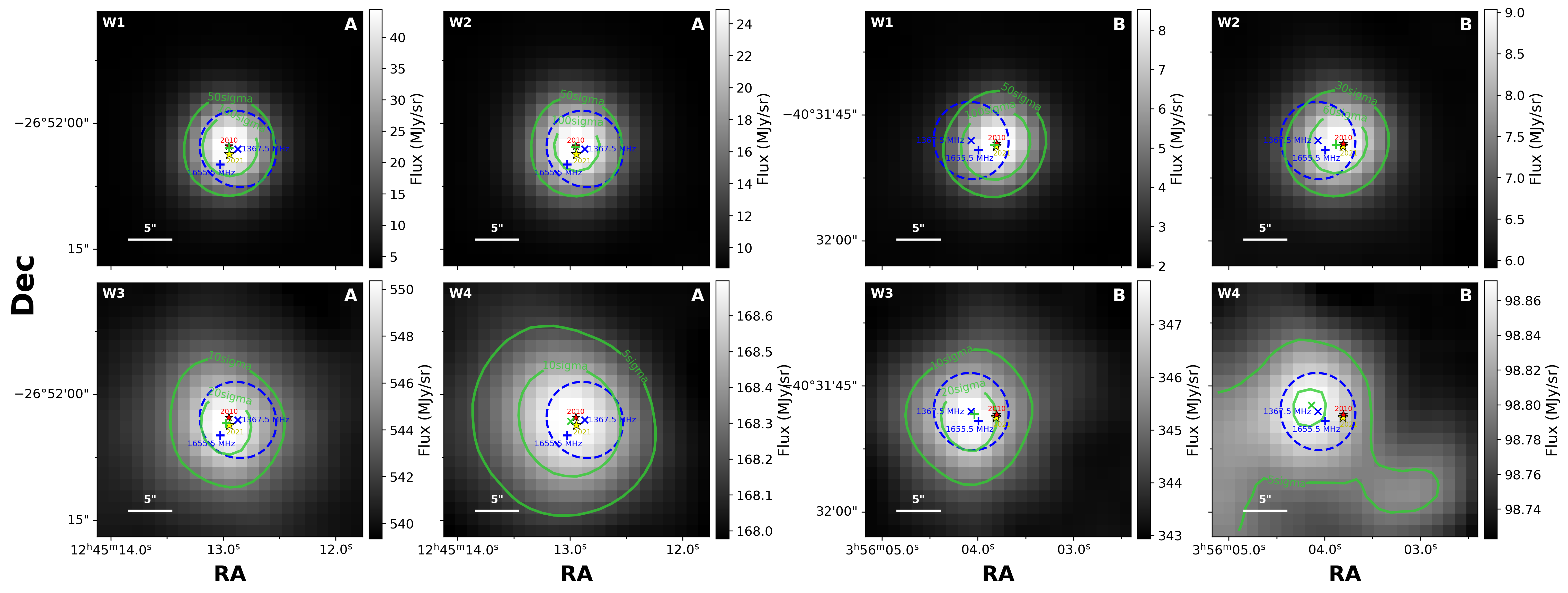}
    \caption{ALLWISE infrared images of radio source candidates A and B across four bands. Each subplot has a field of view of $30\arcsec \times 30\arcsec$, with a $5\arcsec$ scalebar provided for reference. Green contours show surface brightness at multiple $\sigma$ levels above the background, where $\sigma$ is the per-pixel background noise (RMS). Blue dashed ellipses and crosses mark the positions of the 1367.5~MHz radio sources, and the blue plus signs (+) mark the 1655.5~MHz radio sources. The red star and the underlying yellow star indicate the \emph{Gaia} stellar positions corrected to the ALLWISE epoch and the RACS epoch (J2021), respectively. Green crosses ($\times$) and plus signs (+) mark the infrared centroids. Candidate B's $W1$ and $W2$ centroids are offset from the radio source, whereas its $W3$ and $W4$ centroids are consistent with the radio-source position, suggesting the presence of a background infrared source. Candidate~A appears to exhibit a greater degree of positional blending than candidate~B.}
    \label{fig:AB_WISE}
\end{figure*}

Using the ALLWISE centroid detection method described in Section~\ref{sec:section_2}, we obtained the MIR centroids for candidates A and B. The $W1$-, $W2$-, and $W3$-band centroids were measured with \textsc{SEP}, while the $W4$ centroids were measured with \texttt{scipy.ndimage.center\_of\_mass}. These centroids were then compared with the radio counterparts measured at 1367.5~MHz and 1655.5~MHz, which provide superior angular resolution. These WISE centroids were also compared with the stellar positions propagated to the ALLWISE epochs using \emph{Gaia} DR3 proper motions, accounting for the time offset between the ALLWISE observations and the radio measurement epochs. Figure~\ref{fig:AB_WISE} shows the locations of the stars, radio sources, and MIR centroids, and Table~\ref{tab:AB_WISE} lists the measured offsets among the \emph{Gaia} positions, radio positions, and WISE centroids for candidates A and B.

\begin{table*}
\centering
\caption{ALLWISE emission-centroid positions in the $W1$--$W4$ bands, centroiding uncertainties, and angular offsets (with uncertainties) relative to the stellar positions at the ALLWISE epoch and to the radio positions at 1367.5~MHz and 1655.5~MHz for candidates A and B.}
\label{tab:AB_WISE}
\begin{tabular}{cccccccc}
\hline
Candidate & Band & \begin{tabular}{c}RA \\ (h m s)\end{tabular} & \begin{tabular}{c}Dec \\ (d m s)\end{tabular} & \begin{tabular}{c}Centroiding \\ Uncertainty \\ ($\arcsec$)\end{tabular} & \begin{tabular}{c}Offset to star \\ ($\arcsec$)\end{tabular} & \begin{tabular}{c}Offset to 1367.5 MHz \\ ($\arcsec$)\end{tabular} & \begin{tabular}{c}Offset to 1655.5 MHz \\ ($\arcsec$)\end{tabular} \\
\hline
 & $W1$ & 12 45 12.945 & $-26\ 52\ 02.97$ & 0.08 & $0.18 \pm 0.08$ & $1.09 \pm 0.26$ & $2.23 \pm 0.96$ \\
A & $W2$ & 12 45 12.951 & $-26\ 52\ 02.85$ & 0.14 & $0.11 \pm 0.14$ & $1.19 \pm 0.28$ & $2.30 \pm 0.97$ \\
(J1245-2652) & $W3$ & 12 45 12.971 & $-26\ 52\ 03.51$ & 0.17 & $0.81 \pm 0.17$ & $1.49 \pm 0.30$ & $1.59 \pm 0.98$ \\
  & $W4$ & 12 45 12.991 & $-26\ 52\ 03.25$ & 1.10 & $0.78 \pm 1.10$ & $1.71 \pm 1.13$ & $1.75 \pm 1.48$ \\
\cline{2-8}
 & Weighted mean &  &  &  & $0.26 \pm 0.13$ & $1.25 \pm 0.16$ & $2.01 \pm 0.52$ \\
\hline
 & $W1$ & 03 56 03.831 & $-40\ 31\ 48.50$ & 0.21 & $0.31 \pm 0.21$ & $2.81 \pm 0.32$ & $2.01 \pm 0.47$ \\
B & $W2$ & 03 56 03.886 & $-40\ 31\ 48.49$ & 0.32 & $0.91 \pm 0.32$ & $2.20 \pm 0.40$ & $1.43 \pm 0.51$ \\
(J0356-4031) & $W3$ & 03 56 04.042 & $-40\ 31\ 48.33$ & 0.25 & $2.68 \pm 0.25$ & $0.49 \pm 0.36$ & $0.94 \pm 0.40$ \\
 & $W4$ & 03 56 04.140 & $-40\ 31\ 47.28$ & 2.91 & $3.95 \pm 2.91$ & $1.04 \pm 2.92$ & $2.45 \pm 2.93$ \\
\cline{2-8}
 & Weighted mean &  &  &  & $1.22 \pm 0.61$ & $1.87 \pm 0.58$ & $1.41 \pm 0.27$ \\
\hline
\end{tabular}
\vspace{0.3em}
\begin{flushleft}
\footnotesize \textbf{Notes.} Throughout this paper, ``Offset to the star'' represents the angular separation between each MIR centroid and the corresponding \emph{Gaia} position propagated to the corresponding ALLWISE epoch. This convention, featuring the red star as a reference marker, applies specifically to all WISE analysis maps (Figures \ref{fig:AB_WISE}, \ref{fig:CDEF_WISE}, and \ref{fig:HIJ_WISE}) throughout this paper. The ``Weighted mean'' row gives the inverse-variance weighted average of the $W1$--$W4$ offsets in each column; the uncertainty is scaled by $\sqrt{\chi^2/\mathrm{dof}}$ where the band-to-band scatter exceeds the formal errors.
\end{flushleft}
\end{table*}

After accounting for proper motion, candidate A shows a ${\sim 3.3\arcsec}$ offset between the \emph{Gaia} position of the star propagated to J2020.8 and the 887.5~MHz radio source. This is smaller than the 4.9$\arcsec$ offset reported in \citet*{2024RNAAS...8..145R}, which did not correct for the star's proper motion. At higher frequencies with improved angular resolution (1367.5~MHz and 1655.5~MHz), the radio-to-stellar offsets decrease to $\sim1.2\arcsec$ and $\sim1.6\arcsec$ (Table \ref{tab:AB_radio}). The $W1$ and $W2$ centroids lie only $\sim0.2\arcsec$ and $\sim0.1\arcsec$ from candidate star A's position propagated to J2010.5 (Table \ref{tab:AB_WISE}), far closer than the corresponding radio offsets ($\sim1.2\arcsec$ at 1367.5~MHz, $\sim2.3\arcsec$ at 1655.5~MHz). The $W3$ and $W4$ centroid-to-stellar offsets remain small ($\sim0.8\arcsec$), again below the radio offsets. The weighted-mean stellar-to-centroid offset across all four bands for candidate A is $0.26 \pm 0.13\arcsec$ (Table~\ref{tab:AB_WISE}), consistent with the star and substantially smaller than the weighted mean radio-to-stellar offset of $1.27 \pm 0.34\arcsec$ (Table~\ref{tab:AB_radio}).

For candidate B, while the stellar-radio offset at 887.5~MHz is only $\sim0.5\arcsec$, observations with smaller beam sizes at 1367.5~MHz and 1655.5~MHz reveal significantly larger radio-to-stellar offsets exceeding $2\arcsec$ (Table \ref{tab:AB_radio}). Significantly, the WISE $W3$ centroid is substantially displaced from candidate star B's position by $\sim2.7\arcsec$ but is only $\sim0.5\arcsec$ from the 1367.5~MHz radio-source position (Table~\ref{tab:AB_WISE}). In the $W4$ band, which has lower angular resolution than $W3$, the centroid is almost $4\arcsec$ from candidate star B but only $\sim1\arcsec$ from the 1367.5~MHz radio source. In contrast, the $W1$ and $W2$ centroids lie only $\sim0.3\arcsec$ and $\sim0.9\arcsec$ from the star, far closer to the star than to either radio counterpart. The corresponding weighted-mean radio-to-stellar offset for candidate~B is $2.64 \pm 0.36\arcsec$ (Table~\ref{tab:AB_radio}).

\subsection{Astrometric diagnostics for candidates C, D, E, and F}
\label{sec:section_3_2}

\begin{figure*}
    \centering
    \includegraphics[width=1\linewidth]{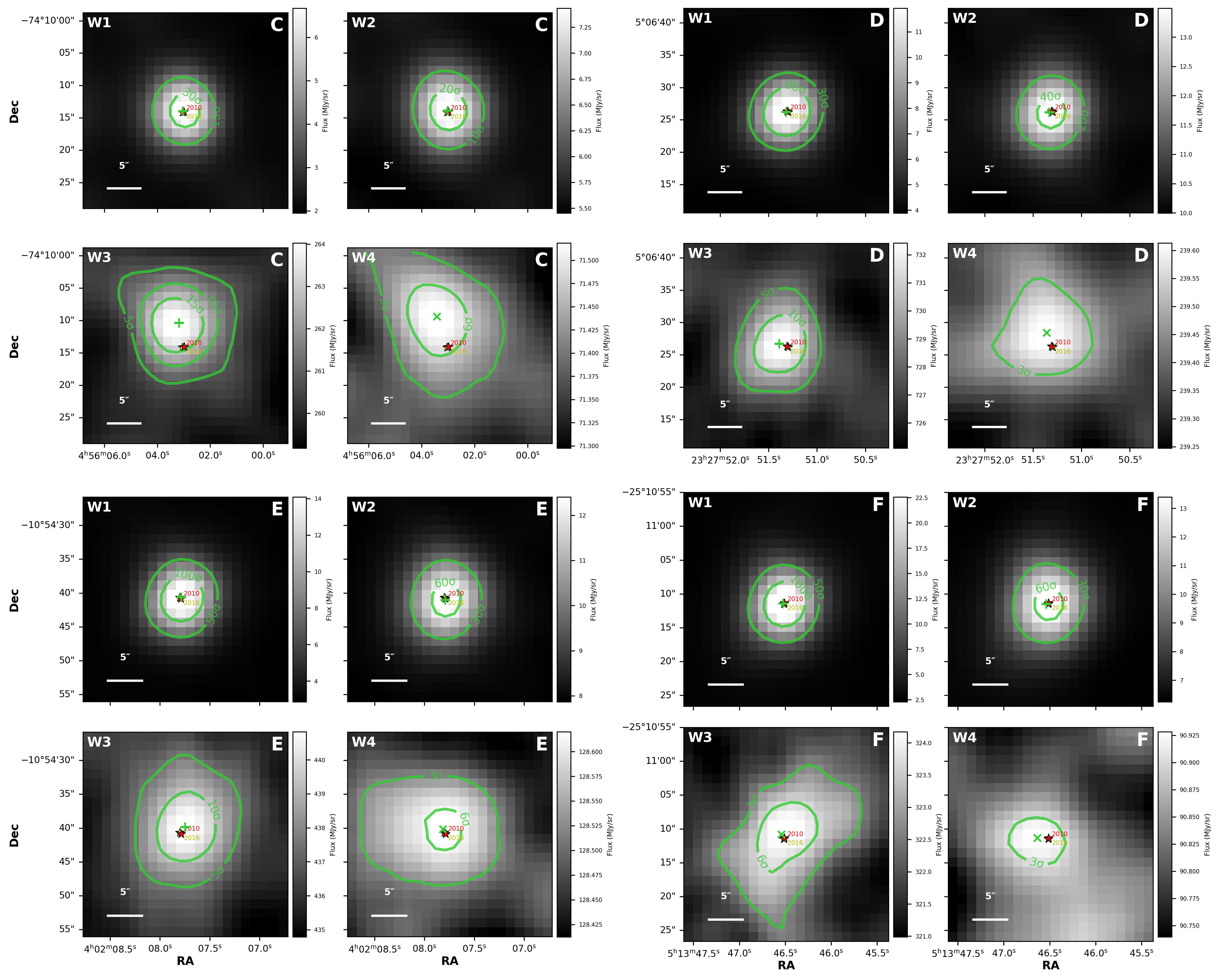}
    \caption{ALLWISE infrared images of the radio-undetected candidates C, D, E, and F across four bands. Each subplot has a field of view of $30\arcsec \times 30\arcsec$, with a $5\arcsec$ scalebar provided for reference. Green contours show surface brightness at multiple $\sigma$ levels above the background, where $\sigma$ is the per-pixel background noise (RMS), and the green crosses ($\times$) and plus signs (+) mark the MIR centroids. The red star and the underlying yellow star indicate the \emph{Gaia} stellar positions corrected to the ALLWISE epoch and the \emph{Gaia} DR3 epoch (2016.0). For the four sources, candidate~C appears to have positional offsets between different WISE bands similar to the pattern of candidate~B, while the MIR centroids and star positions in candidates~D, E, and F appear more blended.}
    \label{fig:CDEF_WISE}
\end{figure*}

We conducted the centroiding analysis for all four \citetalias{2024MNRAS.531..695S} candidate stars with no associated radio sources: C, D, E, and F. All $W1$ and $W2$ centroids for the four cases were obtained using \textsc{SEP}, while all $W4$ images were fitted with \textsc{SciPy}. For the $W3$ band, candidates C, D, and E were successfully fitted with \textsc{SEP}, but the $W3$ emission from candidate F is elongated and could not be fitted by \textsc{SEP}, so we used \textsc{SciPy} for its centroid. The astrometric results for these candidates are shown in Figure~\ref{fig:CDEF_WISE} and Table~\ref{tab:CDEF_WISE}.

\begin{table*}
\centering
\caption{ALLWISE MIR centroid positions, centroiding uncertainties, and offsets (with uncertainties) relative to the propagated \emph{Gaia} positions at the ALLWISE epoch for candidates C, D, E, and F.}
\label{tab:CDEF_WISE}
\begin{tabular}{cccccc}
\hline
Candidate & Band & \begin{tabular}{c}RA \\ (h m s)\end{tabular} & \begin{tabular}{c}Dec \\ (d m s)\end{tabular} & \begin{tabular}{c}Centroiding \\ Uncertainty \\ ($\arcsec$)\end{tabular} & \begin{tabular}{c}Offset to \\ the star \\ ($\arcsec$)\end{tabular} \\
\hline
& $W1$ & 04 56 03.036 & $-74\ 10\ 13.95$ & 0.25 & $0.20 \pm 0.25$ \\
C & $W2$ & 04 56 03.025 & $-74\ 10\ 13.87$ & 0.40 & $0.22 \pm 0.40$ \\
(J0456-7410) & $W3$ & 04 56 03.184 & $-74\ 10\ 10.36$ & 0.27 & $3.77 \pm 0.27$ \\
 & $W4$ & 04 56 03.430 & $-74\ 10\ 09.39$ & 2.43 & $4.99 \pm 2.43$ \\
\hline
 & $W1$ & 23 27 51.319 & $+05\ 06\ 26.17$ & 0.29 & $0.21 \pm 0.29$ \\
D & $W2$ & 23 27 51.332 & $+05\ 06\ 26.16$ & 0.30 & $0.40 \pm 0.30$ \\
(J2327+0506) & $W3$ & 23 27 51.389 & $+05\ 06\ 26.69$ & 0.22 & $1.31 \pm 0.22$ \\
 & $W4$ & 23 27 51.357 & $+05\ 06\ 28.39$ & 1.80 & $2.26 \pm 1.80$ \\
\hline
& $W1$ & 04 02 07.783 & $-10\ 54\ 40.66$ & 0.16 & $0.23 \pm 0.16$ \\
E & $W2$ & 04 02 07.789 & $-10\ 54\ 41.06$ & 0.26 & $0.18 \pm 0.26$ \\
(J0402-1054) & $W3$ & 04 02 07.747 & $-10\ 54\ 39.89$ & 0.24 & $1.15 \pm 0.24$ \\
 & $W4$ & 04 02 07.810 & $-10\ 54\ 40.19$ & 2.26 & $0.81 \pm 2.26$ \\
\hline
 & $W1$ & 05 13 46.524 & $-25\ 11\ 11.44$ & 0.12 & $0.24 \pm 0.12$ \\
 F & $W2$ & 05 13 46.536 & $-25\ 11\ 11.51$ & 0.20 & $0.41 \pm 0.20$ \\
 (J0513-2511)& $W3$ & 05 13 46.539 & $-25\ 11\ 10.82$ & 0.34 & $0.70 \pm 0.34$ \\
 & $W4$ & 05 13 46.632 & $-25\ 11\ 11.37$ & 2.85 & $1.69 \pm 2.85$ \\
\hline
\end{tabular}
\end{table*}

Candidate C (\emph{Gaia} DR3 4649396037451459712) exhibits a positional offset between its emission in the $W1$ and $W2$ bands and that in the $W3$ and $W4$ bands. As shown in Figure \ref{fig:CDEF_WISE}, the $W3$ and $W4$ emission centroids of candidate C are offset by $\sim3.8\arcsec$ and $\sim5.0\arcsec$, respectively, from the \emph{Gaia} position propagated to J2010.3. In contrast, the $W1$ and $W2$ centroid offsets from the \emph{Gaia} position propagated to J2010.8 are both only $\sim0.2\arcsec$.

For candidates D, E, and F, the $W1$ and $W2$ centroid-to-stellar offsets are all below $0.5\arcsec$, and their $W3$ and $W4$ offsets are also substantially smaller than those of candidate~C (Table~\ref{tab:CDEF_WISE}). The $W3$ centroid is offset from the star by $\sim1.3\arcsec$ for candidate~D (\emph{Gaia} DR3 2660349163149053824), $\sim1.1\arcsec$ for candidate~E (\emph{Gaia} DR3 3190232820489766656), and only $\sim0.70\arcsec$ for candidate~F (\emph{Gaia} DR3 2956570141274256512); candidate~E's $W3$ detection is also the most significant of the three, at $>10\sigma$. In the $W4$ band, candidates D and F show offsets of $\sim2\arcsec$, but their $W4$ detections reach only $3$--$5\sigma$ above extended background structure, making these positional offsets poorly constrained. None of the three candidates shows an offset pattern comparable in scale to that of candidate~C.

\subsection{Astrometric diagnostics for candidates H, I, and J}
\label{sec:section_3_3}

\begin{figure*}
    \centering
    \includegraphics[width=1.08\linewidth]{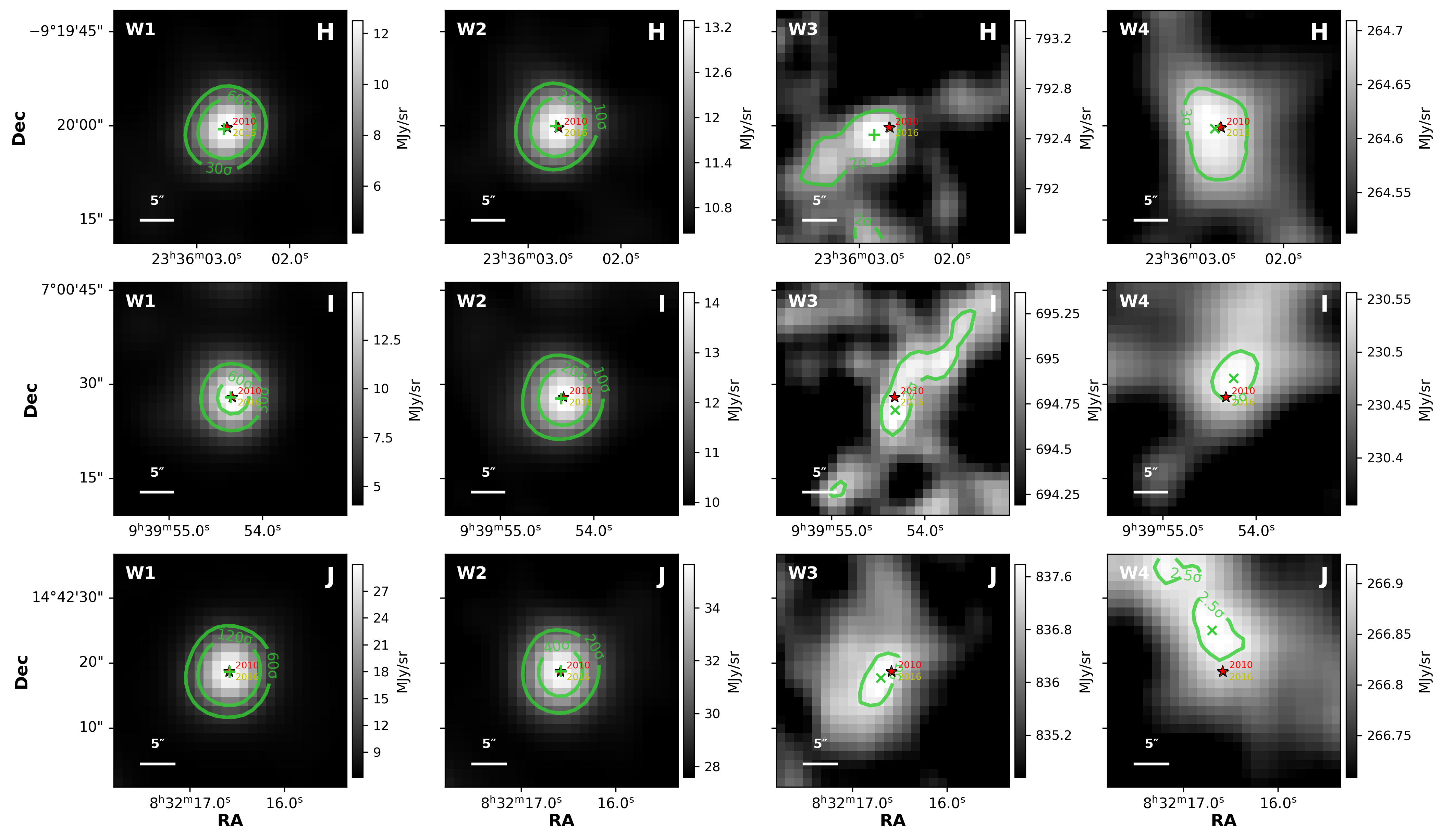}
    \caption{ALLWISE infrared images of candidate stars H, I, and J across four bands. Each subplot has a field of view of $36\arcsec \times 36\arcsec$, with a $5\arcsec$ scalebar provided for reference. Green contours show surface brightness at multiple $\sigma$ levels above the background, where $\sigma$ is the per-pixel background noise (RMS), and the green crosses ($\times$) and plus signs (+) mark the MIR centroids. The red star and the underlying yellow star indicate the \emph{Gaia} stellar positions corrected to the ALLWISE epoch and the \emph{Gaia} DR3 epoch (2016.0). Three candidates generally exhibit lower $W3$ and $W4$ SNRs compared to A--F. Among these, only the $W4$ emission for candidate~H appears positionally aligned; all other $W3$ and $W4$ detections show apparent offsets, with the $W3$ flux for candidate~I notably displaying an extended strip.}
    \label{fig:HIJ_WISE}
\end{figure*}

We present the astrometric analysis of the three new Dyson Sphere candidates in \citetalias{2026arXiv260725701K}. None of these sources has an associated radio or X-ray counterpart based on our VizieR cross-match. The corresponding ALLWISE images and centroid measurements are shown in Figure~\ref{fig:HIJ_WISE} and Table~\ref{tab:HIJ_WISE}.

In the $W1$ and $W2$ bands, three candidates were detected using \textsc{SEP}, and their MIR centroids show offsets below $1\arcsec$. In the $W3$ and $W4$ bands, however, the MIR centroids—except for the $W3$ centroid of candidate H—were measured with \textsc{SciPy}, as the image quality in these bands is too poor for reliable \textsc{SEP} detections. The $W3$ and $W4$ SNRs of candidates H, I, and J are generally low; only the $W3$ image of candidate J reaches a detection significance above $5\sigma$.

Candidate H (\emph{Gaia} DR3 2437221214075471744) shows a $\sim2\sigma$ $W3$ detection with a centroid displaced by $\sim2.6\arcsec$ from the \emph{Gaia} position of the star propagated to J2010.4. In the $W4$ band, the source appears elongated, showing a centroid-to-stellar offset of $\sim1\arcsec$. The detailed spatial structure of this elongated source in the $W4$ band is obscured due to the low image quality.

Similarly, the $W3$ and $W4$ images of candidates I and J exhibit elongated brightness profiles. Candidate I (\emph{Gaia} DR3 3854090071297359616) appears as an extended stripe in $W3$, possibly arising from blended emission associated with multiple nearby sources, and is detected at only $\sim2\sigma$. In $W4$, the source exhibits a smoother morphology with a significance of $\sim3\sigma$. For $W3$, we derived the centroid using the southeastern peak emission, which is closest to the \emph{Gaia} position of candidate star I propagated to J2010.4. The resulting $W3$ and $W4$ centroids lie more than $2\arcsec$ and $3\arcsec$ from the J2010.4 \emph{Gaia} position, respectively. These centroids are oriented in nearly opposite directions, differing by over $5\arcsec$. The $W4$ centroid is offset by $\sim3.2\arcsec$ from candidate I, showing a pattern similar to that seen for candidate C.

Candidate J (\emph{Gaia} DR3 651765552072217216) also shows an offset between the $W3$ and $W4$ centroids. Its centroid-to-stellar offsets are $\sim2\arcsec$ in $W3$ and $\sim6\arcsec$ in $W4$, with opposite directions—the $W3$ and $W4$ centroids of candidate J are separated by more than $7\arcsec$, appearing as if $W3$ and $W4$ are dominated by two different MIR sources.


\begin{table*}
\centering
\caption{ALLWISE MIR centroid positions, centroiding uncertainties, and offsets (with uncertainties) relative to the propagated \emph{Gaia} positions at the ALLWISE epoch for candidates H, I, and J.}
\label{tab:HIJ_WISE}
\begin{tabular}{cccccc}
\hline
Candidate & Band & \begin{tabular}{c}RA \\ (h m s)\end{tabular} & \begin{tabular}{c}Dec \\ (d m s)\end{tabular} & \begin{tabular}{c}Centroiding \\ Uncertainty \\ ($\arcsec$)\end{tabular} & \begin{tabular}{c}Offset to \\ the star \\ ($\arcsec$)\end{tabular} \\
\hline
 & $W1$ & 23 36 02.707 & $-09\ 20\ 00.560$ & 0.19 & $0.55 \pm 0.19$ \\
H & $W2$ & 23 36 02.694 & $-09\ 20\ 00.066$ & 0.34 & $0.38 \pm 0.34$ \\
(J2336-0920) & $W3$ & 23 36 02.834 & $-09\ 20\ 01.461$ & 0.53 & $2.63 \pm 0.53$ \\
 & $W4$ & 23 36 02.733 & $-09\ 20\ 00.496$ & 2.37 & $0.89 \pm 2.37$ \\
\hline
 & $W1$ & 09 39 54.337 & $+07\ 00\ 27.928$ & 0.16 & $0.23 \pm 0.16$ \\
I & $W2$ & 09 39 54.344 & $+07\ 00\ 27.746$ & 0.27 & $0.39 \pm 0.27$ \\
(J0939+0700) & $W3$ & 09 39 54.315 & $+07\ 00\ 25.856$ & 0.62 & $2.10 \pm 0.62$ \\
 & $W4$ & 09 39 54.239 & $+07\ 00\ 30.934$ & 2.15 & $3.23 \pm 2.15$ \\
\hline
 & $W1$ & 08 32 16.575 & $+14\ 42\ 18.643$ & 0.11 & $0.11 \pm 0.11$ \\
J & $W2$ & 08 32 16.575 & $+14\ 42\ 18.751$ & 0.20 & $0.09 \pm 0.20$ \\
(J0832+1442) & $W3$ & 08 32 16.694 & $+14\ 42\ 17.700$ & 0.33 & $1.94 \pm 0.33$ \\
 & $W4$ & 08 32 16.694 & $+14\ 42\ 24.985$ & 2.17 & $6.46 \pm 2.17$ \\
\hline
\end{tabular}
\end{table*}

\subsection{Comparison with the Photocentre for the S24 Candidate Stars}
\label{sec:section_3_4}

Candidates A and B in Subsection \ref{sec:section_3_1} and C, D, E, and F in Subsection \ref{sec:section_3_2} were analysed for photocentre offsets in \citetalias{2024MNRAS.531..695S}. Although our centroiding methods are not identical to those used in \citetalias{2024MNRAS.531..695S}, our results are broadly consistent with the values presented in Table 7 of \citetalias{2024MNRAS.531..695S}. Our calculated $W1$/$W2$ total offsets remain below $0.5\arcsec$. For candidates B and C, the $W1$/$W3$ total offsets of $2.43\arcsec$ and $3.64\arcsec$ align with the significant displacements reported for those cases in \citetalias{2024MNRAS.531..695S}. Table \ref{tab:centroid_comparison} presents a comparison between the results from \citetalias{2024MNRAS.531..695S} and this work.

\begin{table*}
\centering
\caption{Comparison of photocentre offsets (arcsec) between \citetalias{2024MNRAS.531..695S} and this work.}
\label{tab:centroid_comparison}
\begin{tabular}{ccccccc}
\hline
\multirow{2}{*}{Candidate} &
\multicolumn{2}{c}{\makecell{$W1$/$W2$ offset\\(RA/Dec; $\arcsec$)}} &
\multicolumn{2}{c}{\makecell{$W1$/$W3$ offset\\(RA/Dec; $\arcsec$)}} &
\multicolumn{1}{c}{\makecell{$W2$/$W3$ offset\\(RA/Dec; $\arcsec$)}} \\
\cmidrule(lr){2-3}
\cmidrule(lr){4-5}
\cmidrule(lr){6-6}
&
S24 &
This Work &
S24 &
This Work &
This Work \\
\hline
A (J1245-2652) & $-0.25 / -0.01$ & $0.00 / 0.01$  & $-0.03 / 0.33$  & $0.33 / -0.65$ & \multicolumn{2}{c}{$0.33 / -0.66$} \\
B (J0356-4031)& $0.40 / 0.31$   & $0.63 / 0.03$  & $3.21 / 0.06$   & $2.43 / 0.15$  & \multicolumn{2}{c}{$1.80 / 0.12$} \\
C (J0456-7410) & $0.25 / -0.32$  & $-0.05 / 0.08$ & $1.52 / -3.68$  & $0.77 / 3.56$  & \multicolumn{2}{c}{$0.81 / 3.48$} \\
D (J2327+0506)& $-0.31 / -0.12$ & $0.19 / -0.01$ & $0.60 / -0.09$  & $1.05 / 0.52$  & \multicolumn{2}{c}{$0.85 / 0.53$} \\
E (J0402-1054)& $-0.09 / 0.48$  & $0.10 / -0.40$ & $-1.15 / -0.38$ & $-0.52 / 0.73$ & \multicolumn{2}{c}{$-0.62 / 1.13$} \\
F (J0513-2511)& $0.03 / 0.10$   & $0.19 / -0.08$ & $-1.04 / 0.79$  & $-0.08 / 1.18$ & \multicolumn{2}{c}{$-0.27 / 1.26$} \\
\hline
\end{tabular}
\end{table*}

For the remaining candidates (A, D, E, and F), $W1$/$W3$ positional offsets range from $0.7\arcsec$ to $1.2\arcsec$. These do not deviate from the broad distributions suggested in \citet{2017AJ....153..165T} ($\sigma \approx 5\arcsec$), where band-to-band centroiding was applied to investigate warm dust around M dwarfs. While utilising precise \emph{Gaia} DR3 positions in this work provides a better stellar reference, it does not inherently mitigate the degeneracy caused by such wide statistical ranges. To better examine the potential background contaminants of the candidate stars, we selected sets of SNR-matched control stars for a comparative analysis, as detailed in Section \ref{sec:section_3_5}.

\subsection{ALLWISE Image Centroid Analysis of Ordinary M-Dwarf Control Stars}
\label{sec:section_3_5}

To evaluate the statistical significance of our astrometric offsets, we constructed nine distinct control groups—one corresponding to each of the investigated candidates from A to J, except for candidate G, as it was previously identified as a clear case of AGN contamination \citep*{2025MNRAS.538L..56R}.

We first constructed an initial parent sample of ordinary M-dwarfs by selecting sources from \emph{Gaia} DR3 with a Renormalised Unit Weight Error ($\mathrm{RUWE}$) $< 1.4$ located within $300\text{~pc}$. To specifically isolate the M-dwarf population, we applied multiple astrophysical constraints, including an absolute magnitude cut of $M_G > 7.7$, an effective temperature threshold of $T_{\mathrm{eff}} < 4000\text{~K}$, and a colour cut of $G_{\mathrm{BP}}-G_{\mathrm{RP}} \ge 1.84$ \citep{2021A&A...653A..98M}. Following cross-matching with the 2MASS and ALLWISE catalogues, this initial selection yielded approximately 2 million sources.

Subsequently, we applied a series of stringent ALLWISE photometric quality filters to ensure the reliability of the MIR detections. We required a signal-to-noise ratio $\mathrm{SNR} > 3$ in both the $W3$ and $W4$ bands, a criterion that excluded over 99 per cent of the sources. Additionally, we imposed profile-fit $\chi^2$ limits ($\mathrm{w3rchi2} < 1.05$ and $\mathrm{w4rchi2} < 1.3$), mandated the absence of saturation, and required at least one valid profile-fit measurement in both the $W3$ and $W4$ bands. Collectively, these requirements reduced the parent population to a final baseline sample of 3,582 high-quality ordinary M-dwarfs (see Table~\ref{tab:selection_criteria}).

\begin{table*}
\centering
\caption{Summary of the selection criteria used to construct the M-dwarf control sample.}
\label{tab:selection_criteria}
\begin{tabular}{>{\centering\arraybackslash}m{3.6cm} p{9cm} >{\centering\arraybackslash}m{2cm}}
\hline
Step & Filter criterion & Source count \\
\hline
1 Initial Sample Selection
& \emph{Gaia} DR3 RUWE $<1.4$ and distance $<300$~pc
& 6,346,475 \\
\hline
\multirow{3}{3.6cm}{\centering 2 M-dwarf Selection}
& Absolute magnitude $M_G > 7.7$ (pass rate: 85\%) & --- \\
& 2MASS and ALLWISE best-neighbour match (pass rate: 61\%) & --- \\
& $T_{\mathrm{eff}} < 4000$~K and $\mathrm{BP}-\mathrm{RP} \geq 1.84$ (pass rate: 61\%) & 2,016,459 \\
\hline
\multirow{4}{3.6cm}{\centering 3 Photometric Quality Selection}
& Both $W3$ and $W4$ S/N $> 3$ (pass rate: 1\%) & \multirow{4}{*}{3,582} \\
& Profile-fit $\chi^2$: $\mathrm{w3rchi2} < 1.05$ and $\mathrm{w4rchi2} < 1.3$ (pass rate: 88\%) & \\
& No saturation: $\mathrm{w3sat}=0$ and $\mathrm{w4sat}=0$ (pass rate: 97\%) & \\
& Number of measurements: $\mathrm{w3nm}>0$ and $\mathrm{w4nm}>0$ (pass rate: 30\%) & \\
\hline
\end{tabular}
\end{table*}

To establish a reliable astrometric baseline, we matched stars based on their \emph{Gaia} DR3 $G$-band magnitudes, $W1$ SNR, and $W2$ SNR. This selection relies on the premise that at these shorter wavelengths, corresponding to the Rayleigh--Jeans tail, the SEDs of potential Dyson Sphere candidates should remain similar to those of ordinary M-dwarfs. This matching criterion initially yielded 406 stars selected from the parent sample of 3,582. We then performed a direct visual inspection to ensure clear profile structures in the $W3$ and $W4$ bands, filtering the sample down to 155 final control stars across the nine groups (see Table~\ref{tab:control_samples}).

\begin{table*}
\centering
\caption{The candidate star properties and the corresponding control sample selection. The left section lists properties for the candidate stars, while the right section defines the selection criteria and resulting sample sizes for the control groups. SNR values are taken from the ALLWISE catalogue and $G$-band Vega magnitudes are from \emph{Gaia} DR3. $N_{\text{sel}}$ shows the number of control stars that passed the SNR and $G$-magnitude selection, while $N_{\text{fit}}$ denotes the number of control stars that passed the visual check and were therefore used for the Rayleigh distribution analysis.}
\label{tab:control_samples}
\setlength{\tabcolsep}{6pt}
\begin{tabular}{c ccc ccccc}
\hline
\multirow{2}{*}{Candidate} &
\multicolumn{3}{c}{Candidate Star} &
\multicolumn{5}{c}{Control Sample (M-dwarfs)} \\
\cmidrule(lr){2-4}
\cmidrule(lr){5-9}
&
$\mathrm{SNR}_{W1}$ &
$\mathrm{SNR}_{W2}$ &
$G$-mag &
$W1$ range &
$W2$ range &
$G$-mag range &
$N_{\mathrm{sel}}$ &
$N_{\mathrm{fit}}$ \\
\midrule
A (J1245$-$2652) & 45.8 & 48.2 & 15.996 & 44.8--46.8 & 47.2--49.2 & 15.75--16.25 & 25 & 15 \\
B (J0356$-$4031) & 43.7 & 39.5 & 17.713 & 41.7--45.7 & 37.5--41.5 & 17.20--18.20 & 58 & 20 \\
C (J0456$-$7410) & 41.8 & 33.3 & 18.393 & 39.8--43.8 & 31.3--35.3 & 17.90--18.90 & 28 & 16 \\
D (J2327$+$0506) & 41.9 & 31.4 & 17.662 & 39.9--43.9 & 29.4--33.4 & 17.20--18.20 & 29 & 13 \\
E (J0402$-$1054) & 43.3 & 38.4 & 17.008 & 41.3--45.3 & 36.4--40.4 & 16.50--17.50 & 94 & 25 \\
F (J0513$-$2511) & 44.0 & 42.4 & 16.330 & 42.0--46.0 & 40.4--44.4 & 15.80--16.80 & 95 & 37 \\
H (J2336$-$0920) & 41.6 & 29.5 & 17.312 & 39.6--43.6 & 27.5--31.5 & 16.30--18.30 & 23 & 9  \\
I (J0939$+$0700) & 44.4 & 34.9 & 17.402 & 42.4--46.4 & 32.9--36.9 & 16.90--17.90 & 26 & 9  \\
J (J0832$-$1442) & 44.7 & 32.6 & 16.133 & 40.0--45.0 & 28.0--38.0 & 15.00--17.00 & 28 & 11 \\
\hline
\end{tabular}
\end{table*}

We then performed a centroid analysis on the 155 stars, following the same procedure introduced in Section~\ref{sec:section_2} and applied in Subsections~\ref{sec:section_3_1}, \ref{sec:section_3_2}, and \ref{sec:section_3_3}. For each star where a centroid was successfully detected, the centroiding process yielded both a positional offset and the corresponding uncertainty. The positional offsets of the control stars, used as the reference for the candidate stars, were then quantitatively analysed using the Rayleigh distribution by employing \texttt{scipy.stats.rayleigh}, with the quantitative results presented in Table~\ref{tab:rayleigh}.

The typical offset scale of the M~dwarfs in each band of each control sample is characterised by the Rayleigh scale parameter $\mathrm{\sigma_R}$. Each candidate's offset $\mathrm{r_{cand}}$ is compared against this scale to determine a Rayleigh survival probability $\mathrm{p_R}$, which represents the chance of a control star showing an offset at least as large. We then recast this probability as a one-sided Gaussian-equivalent significance, $\Phi^{-1}(1-\mathrm{p_R})$, which enables the standard-deviation expression. For all control samples, we also calculate an outlier-insensitive median scale $\mathrm{\sigma_{med}}$ as an independent baseline. This serves as a reliable backup because some datasets, most notably the $W3$ and $W4$ bands, do not strictly follow a Rayleigh distribution. Appendix~\ref{sec:appendix} presents further details of the Rayleigh-distribution analysis beyond the table, including the full centroid-offset distributions for all candidates and bands and an example of the control-star centroiding for candidate~A.

\begin{table*}
\centering
\caption{Rayleigh-distribution analysis of centroid offsets for the candidate control samples (W1--W4). $N$ is the number of valid control stars; $\sigma_R$ is the Rayleigh scale of the control sample, where the Rayleigh distribution peaks at (illustrated in Fig.~\ref{fig:control_offset}); $p_R$ is the Rayleigh survival probability at the candidate offset; and the Gaussian-equivalent significances are the one-sided $\sigma$ values transformed from $p_R$ under the Rayleigh scale $\sigma_R$ and under the outlier-insensitive median scale $\sigma_{\rm med}$, respectively.}
\label{tab:rayleigh}
\begin{tabular}{c l r c c c c}
\hline
\multirow{2}{*}{Candidate} & \multirow{2}{*}{Band} & \multirow{2}{*}{$N$} &
Rayleigh scale & Rayleigh survival & \multicolumn{2}{c}{Gaussian-equivalent significance ($\sigma$)} \\
\cline{6-7}
 & & & $\sigma_R$ (arcsec) & probability $p_R$ & Rayleigh & Median \\
\hline
\multirow{4}{*}{A (J1245$-$2652)} & $W1$ & 15 & $0.17 \pm 0.02$ & $0.585$ & $-0.22 \pm 1.15$ & $+0.11 \pm 1.35$ \\
 & $W2$ & 15 & $0.21 \pm 0.03$ & $0.874$ & $-1.15 \pm 3.06$ & $-1.04 \pm 3.23$ \\
 & $W3$ & 15 & $1.50 \pm 0.36$ & $0.864$ & $-1.10 \pm 0.41$ & $-0.75 \pm 0.45$ \\
 & $W4$ & 9 & $3.29 \pm 0.56$ & $0.972$ & $-1.91 \pm 2.79$ & $-1.89 \pm 2.88$ \\
\hline
\multirow{4}{*}{B (J0356$-$4031)} & $W1$ & 19 & $0.26 \pm 0.06$ & $0.479$ & $+0.05 \pm 2.41$ & $+1.08 \pm 2.76$ \\
 & $W2$ & 19 & $0.28 \pm 0.03$ & $0.006$ & $+2.53 \pm 1.46$ & $+2.35 \pm 1.63$ \\
 & $W3$ & 18 & $2.29 \pm 0.31$ & $0.504$ & $-0.01 \pm 0.30$ & $+0.25 \pm 0.52$ \\
 & $W4$ & 9 & $2.80 \pm 0.39$ & $0.371$ & $+0.33 \pm 2.73$ & $+0.18 \pm 2.77$ \\
\hline
\multirow{4}{*}{C (J0456$-$7410)} & $W1$ & 14 & $0.33 \pm 0.09$ & $0.835$ & $-0.97 \pm 3.25$ & $-0.40 \pm 3.54$ \\
 & $W2$ & 14 & $0.35 \pm 0.06$ & $0.818$ & $-0.91 \pm 3.76$ & $-0.37 \pm 4.13$ \\
 & $W3$ & 12 & $2.59 \pm 0.35$ & $0.347$ & $+0.39 \pm 0.33$ & $+0.76 \pm 0.50$ \\
 & $W4$ & 7 & $2.74 \pm 0.72$ & $0.191$ & $+0.87 \pm 1.99$ & $+1.63 \pm 2.24$ \\
\hline
\multirow{4}{*}{D (J2327$+$0506)} & $W1$ & 13 & $0.33 \pm 0.10$ & $0.817$ & $-0.90 \pm 3.50$ & $+0.45 \pm 4.29$ \\
 & $W2$ & 13 & $0.35 \pm 0.06$ & $0.524$ & $-0.06 \pm 2.59$ & $+0.18 \pm 2.96$ \\
 & $W3$ & 10 & $2.39 \pm 0.36$ & $0.860$ & $-1.08 \pm 0.28$ & $-1.16 \pm 0.35$ \\
 & $W4$ & 5 & $3.30 \pm 0.74$ & $0.791$ & $-0.81 \pm 2.43$ & $-0.67 \pm 2.44$ \\
\hline
\multirow{4}{*}{E (J0402$-$1054)} & $W1$ & 25 & $0.23 \pm 0.03$ & $0.595$ & $-0.24 \pm 2.27$ & $-0.11 \pm 2.35$ \\
 & $W2$ & 25 & $0.30 \pm 0.02$ & $0.830$ & $-0.96 \pm 3.36$ & $-1.12 \pm 3.28$ \\
 & $W3$ & 20 & $2.57 \pm 0.47$ & $0.905$ & $-1.31 \pm 0.32$ & $-0.93 \pm 0.35$ \\
 & $W4$ & 14 & $2.48 \pm 0.17$ & $0.948$ & $-1.62 \pm 3.60$ & $-1.77 \pm 3.50$ \\
\hline
\multirow{4}{*}{F (J0513$-$2511)} & $W1$ & 36 & $0.53 \pm 0.20$ & $0.903$ & $-1.30 \pm 1.36$ & $-0.18 \pm 1.45$ \\
 & $W2$ & 37 & $0.46 \pm 0.14$ & $0.673$ & $-0.45 \pm 1.44$ & $+0.91 \pm 1.73$ \\
 & $W3$ & 28 & $2.28 \pm 0.24$ & $0.954$ & $-1.68 \pm 0.99$ & $-1.45 \pm 1.04$ \\
 & $W4$ & 21 & $2.99 \pm 0.21$ & $0.853$ & $-1.05 \pm 3.51$ & $-1.21 \pm 3.38$ \\
\hline
\multirow{4}{*}{H (J2336$-$0920)} & $W1$ & 7 & $0.56 \pm 0.11$ & $0.618$ & $-0.30 \pm 0.79$ & $-0.17 \pm 1.15$ \\
 & $W2$ & 7 & $0.51 \pm 0.07$ & $0.760$ & $-0.71 \pm 2.67$ & $-0.99 \pm 2.69$ \\
 & $W3$ & 5 & $3.54 \pm 0.83$ & $0.759$ & $-0.70 \pm 0.48$ & $-0.71 \pm 0.69$ \\
 & $W4$ & 4 & $2.35 \pm 0.36$ & $0.931$ & $-1.48 \pm 3.71$ & $-1.64 \pm 3.61$ \\
\hline
\multirow{4}{*}{I (J0939$+$0700)} & $W1$ & 7 & $0.16 \pm 0.04$ & $0.367$ & $+0.34 \pm 2.73$ & $+1.64 \pm 3.17$ \\
 & $W2$ & 7 & $0.16 \pm 0.03$ & $0.044$ & $+1.71 \pm 3.30$ & $+2.18 \pm 3.56$ \\
 & $W3$ & 3 & $1.94 \pm 0.34$ & $0.557$ & $-0.14 \pm 0.65$ & $-0.35 \pm 0.73$ \\
 & $W4$ & 6 & $2.53 \pm 0.49$ & $0.442$ & $+0.15 \pm 2.42$ & $+0.23 \pm 2.59$ \\
\hline
\multirow{4}{*}{J (J0832$+$1442)} & $W1$ & 10 & $0.29 \pm 0.08$ & $0.933$ & $-1.50 \pm 2.58$ & $-1.06 \pm 2.80$ \\
 & $W2$ & 10 & $0.38 \pm 0.08$ & $0.973$ & $-1.92 \pm 3.21$ & $-1.79 \pm 3.37$ \\
 & $W3$ & 9 & $1.49 \pm 0.28$ & $0.429$ & $+0.18 \pm 0.50$ & $+0.04 \pm 0.63$ \\
 & $W4$ & 5 & $2.93 \pm 0.26$ & $0.088$ & $+1.35 \pm 1.05$ & $+0.78 \pm 1.04$ \\
\hline
\end{tabular}
\end{table*}

Only the $W2$ band of candidate~B exceeds $2\sigma$ under both scales, with a Gaussian-equivalent significance of $2.53 \pm 1.46\sigma$ (Rayleigh) and $2.35 \pm 1.63\sigma$ (median), supported by a well-behaved control-sample fit (Table~\ref{tab:rayleigh}). The only other cell to reach $2\sigma$ under either scale is the $W2$ band of candidate~I, at $2.18 \pm 3.56\sigma$ under the median scale but only $1.71 \pm 3.30\sigma$ under the Rayleigh scale; with an uncertainty larger than the value itself and just 7 control stars, we regard this as inconclusive rather than indicative of a real offset for candidate~I. We also note that a large raw offset does not by itself imply a significant one. The $W3$ centroid of candidate~C lies $3.77 \pm 0.27\arcsec$ from the star (Table~\ref{tab:CDEF_WISE}), yet corresponds to only $0.39 \pm 0.33\sigma$ (Rayleigh) and $0.76 \pm 0.50\sigma$ (median), since its control stars show a comparably large offset scatter ($\sigma_R = 2.59 \pm 0.35\arcsec$).

\section{Archival Image Diagnostics}
\label{sec:section_4}

To identify potential background contaminants that escape detection in the \emph{Gaia}, 2MASS, and WISE surveys with shallower depth, we utilise deep archival optical and NIR photometric data. In the optical regime, all candidates are covered by the deep DESI Legacy Survey DR10 imaging, which offers superior depth compared to shallower surveys (e.g., SDSS, SkyMapper, and Pan-STARRS). In the NIR regime, coverage varies depending on the specific survey footprints: A, B, E, F, and H are imaged by the VHS survey, while only candidate C is covered by the VMC survey. The remaining three candidates lack VISTA imaging but are instead covered by UKIRT surveys, with D and I observed by UKIDSS and J captured by the UHS. Consequently, we divide the sample into three groups based on their available near-infrared survey coverage:

\begin{itemize}
    \item Candidates A, B, E, F, and H (with VHS coverage);
    \item Candidate C (with VMC coverage);
    \item Candidates D, I, and J (with UKIDSS or UHS coverage).
\end{itemize}

We complement our image inspection by examining the associated VISTA and UKIRT catalogues in the NIR regime, and DESI Legacy DR10 \textsc{Tractor} optical catalogues to find identifiers and existing photometric classifications for sources apparent in the images of each candidate. The VISTA and UKIRT catalogues provide basic stellar/galaxy morphological classifications, and the Legacy \textsc{Tractor} catalogue offers detailed brightness profile types ($\text{PSF}$, $\text{REX}$, $\text{EXP}$, etc.) and properties. The combined optical and NIR approach enables a preliminary diagnosis of the associated sources.

To identify potentially uncatalogued radio sources that might indicate AGN contamination, we also inspected archival radio imaging, including the NRAO VLA Sky Survey (NVSS; \citealt{1998AJ....115.1693C}), the Faint Images of the Radio Sky at Twenty-Centimeters (FIRST; \citealt*{1995ApJ...450..559B}), the VLA Low-frequency Sky Survey Redux (VLSSr; \citealt{2014MNRAS.440..327L}), and the Sydney University Molonglo Sky Survey (SUMSS; \citealt{2003MNRAS.342.1117M}), for sources lacking radio catalogue counterparts. No new radio candidates with significance exceeding $3\sigma$ were identified.

We applied the imaging and photometric inspection method to candidate G as a verification case, utilising its Legacy Survey DR10, VHS and UHS data; the findings are described in Subsection \ref{sec:section_4_1}. The diagnostic results for the three groups, derived from archival image inspections, follow in Subsections \ref{sec:section_4_2}, \ref{sec:section_4_3}, and \ref{sec:section_4_4}.

\subsection{Searching for the AGN associated with candidate G in the optical and NIR regimes}
\label{sec:section_4_1}

To validate the effectiveness of detecting background contaminants using deep archival imaging, we first apply our analysis methodology to the hot DOG associated with candidate G (J2335–0004). This object was recently confirmed as a radio-loud AGN by \citet*{2025MNRAS.538L..56R} using VLBI astrometry, and the ALLWISE positional offset analysis supports its hot DOG properties. We examined the same archival Legacy, VISTA and UKIRT survey datasets used for the nine new candidates to determine if this known contaminant could be identified solely through archival inspection.

\begin{figure}
    \centering
    \includegraphics[width=1\linewidth]{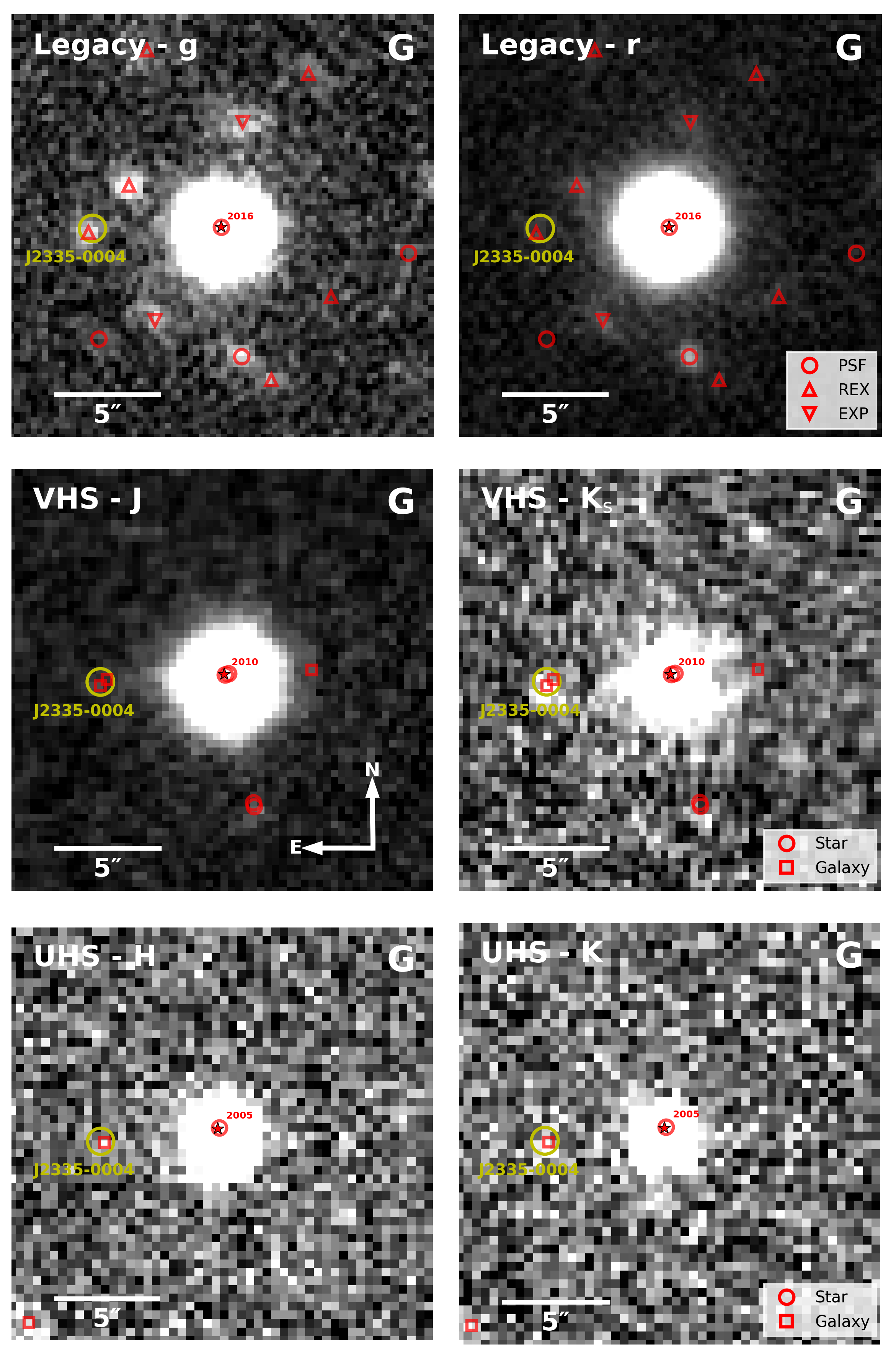}
    \caption{Legacy DR10 $g$- and $r$-band images (upper panels), VHS $J$- and $K_{\rm s}$-band images (middle panels), and UHS $H$- and $K$-band images (lower panels) for candidate G. Each panel shows a single red star marking the \emph{Gaia} position propagated to that survey's epoch (J2016 for Legacy, 2010 for VHS, 2005 for UHS), while the highlighted circles indicate the AGN core position detected by EVN(e-VLBI). In the Legacy $g$-band panels, red circles indicate sources with PSF profiles, while upward- and downward-pointing red triangles mark EXP sources, respectively. In the VHS and UHS panels, red circles indicate stellar objects, and green squares mark the galaxy-like objects. A compass marker in the bottom-right of the middle-left panel (VHS $J$) indicates orientation (N/E) for all subplots.}
    \label{fig:G_image}
\end{figure}

J2335-0004 was effectively detected by the DESI Legacy Survey. The Legacy DR10 Tractor catalogue suggested that this object has an $\text{REX}$ profile, consistent with an extended source (a galaxy). It is notably faint across optical bands, with $g, r, i,$ and $z$ AB magnitudes of 24.43, 24.23, 24.20, and 23.11 mag, respectively. This Legacy object in the optical regime is located $0.28\arcsec$ away from the AGN core position of J2335-0004 (23:35:32.8406, -0:04:25.069) detected through the EVN (e-VLBI)

In the VHS survey, J2335--0004 was reported to be an extended, galaxy-like source, with a $K_{\rm s}$-band AB magnitude of $19.89 \pm 0.26$~mag. Additionally, the VHS DR5 catalogue \citep{2013Msngr.154...35M} reported a double detection at this location: VHS J233532.82-000424.9 and VHS J233532.84-000425.2. Both detections are found to be highly coincident, lying $\sim0.3\arcsec$ away from the AGN core position of J2335-0004. In another NIR survey, this source was catalogued in the UKIDSS DR9 Large Area Survey \citep{2007MNRAS.379.1599L} as ULAS J233532.83-000425.1; it is classified as a galaxy-like object with a $K$-band AB magnitude of $20.09 \pm 0.21$~mag.

We also inspected archival image data from the SkyMapper, Pan-STARRS, and SDSS; however, the faint optical magnitude (as revealed by the Legacy DR10 catalogue) falls below their respective detection thresholds, and we did not detect it. These findings on the AGN companion of candidate G effectively demonstrate that deep public archival surveys can identify background contaminants at minimal cost if their depth and sensitivity are adequate. However, this archival evidence alone is insufficient for a complete MIR contributor diagnosis. A further astrometric analysis, similar to that performed in Section \ref{sec:section_3}, is required to confirm that the MIR emission genuinely originates from the background galaxy revealed in the archival images.

\subsection{Image diagnostics for candidates A, B, E, F, and H}
\label{sec:section_4_2}

The five candidates covered by the VHS survey include A and B, both of which have radio counterparts in the public surveys, and E, F, and H, which show no associated radio emission.

\begin{figure}
    \centering
    \includegraphics[width=1.0\linewidth]{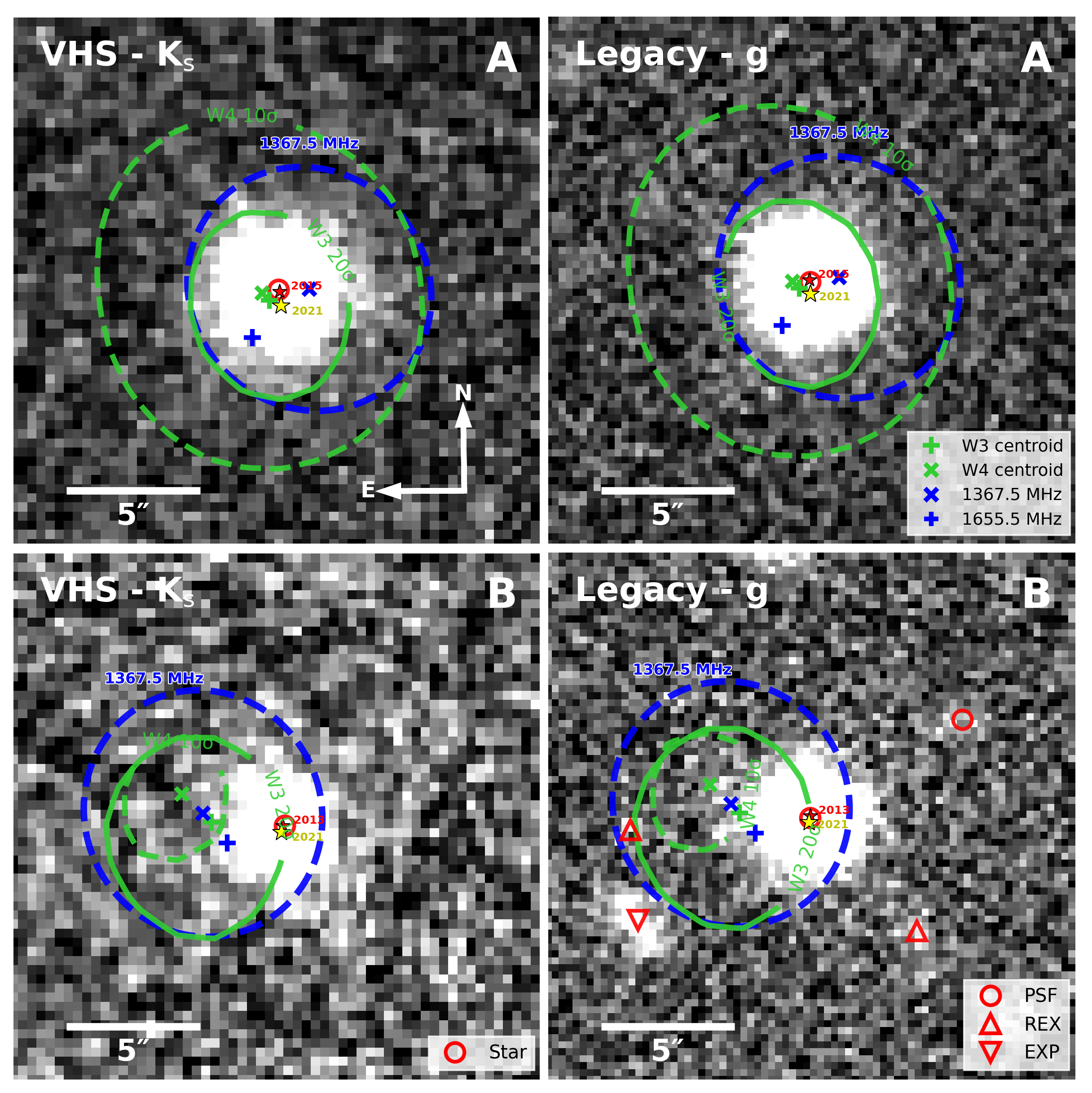}
    \caption{VHS $K_{\rm s}$-band (left panels) and Legacy $g$-band (right panels) images of candidates A and B. Red stars mark the propagated \emph{Gaia} positions at the VHS imaging epochs (2015 for A and 2013 for B), while yellow stars show the propagated \emph{Gaia} positions at the RACS observation epoch (2021). Green plus signs ($+$) and crosses ($\times$) mark the $W3$ and $W4$ centroids, respectively. The corresponding green $W3$ (solid) and $W4$ (dashed) high-flux contours for candidates A and B are also shown, with all data measured in Section \ref{sec:section_3}. The 1367.5~MHz radio source positions are indicated by blue crosses ($\times$), with blue ellipses representing the 1367.5~MHz radio source size as determined by a \textsc{pyBDSF} fit. Blue plus signs (+) mark 1655.5 MHz radio-detected positions. In the VHS $K_{\rm s}$-band panels, red circles indicate stellar objects from the VHS catalogue. In the Legacy $g$-band panels, red circles indicate sources with PSF profiles, while red triangles mark REX sources using upward triangles and EXP sources using downward triangles. North (N) and east (E) directions are indicated by the compass marker in the bottom-right corner of the upper-left panel and apply to all cutouts. While the WISE emissions, the position of the candidate stars, and the radio source position for candidate A are spatially blended, candidate B exhibits alignment between the radio and WISE emissions, both of which are separated from the position of the star.}
    \label{fig:AB_Images}
\end{figure}

Figure~\ref{fig:AB_Images} presents the VHS $K_{\rm s}$-band and Legacy DR10 $g$-band images for candidates A and B, the two sources with confirmed radio counterparts. For candidate A, we find a significant spatial coincidence between the WISE $W3$ and $W4$ centroids, the 1367.5~MHz and 1655.5~MHz radio positions, and the central optical/NIR source brightness profile in Figure~\ref{fig:AB_Images}. The source maintains a consistent, unresolved morphology across all available archival bands (VHS $J$ and $Y$; Legacy $r$, $i$, and $z$). The associated VHS and Legacy \textsc{Tractor} catalogues do not show any companion objects within $10\arcsec$ from candidate A.

For candidate B, as revealed by Section~\ref{sec:section_3_1}, the $1367.5~\mathrm{MHz}$ and $1655.5~\mathrm{MHz}$ radio counterparts, along with the WISE $W3$ and $W4$ centroids, are tightly clustered $\sim 3\arcsec$ to the east of the candidate B. This position does not exhibit significant emission in the VHS $K_{\rm s}$-band image (Figure~\ref{fig:AB_Images}) and is not detected in the other available VHS bands. Furthermore, no close companion source is catalogued at this position in either the VHS or Legacy DR10 Tractor catalogues. However, the Legacy $g$-band image reveals a faint, tail-like structure at this offset location, consistent with the radio and WISE positions, detected with a peak significance of $> 3\sigma$.

\begin{figure}
    \centering
    \includegraphics[width=1.0\linewidth]{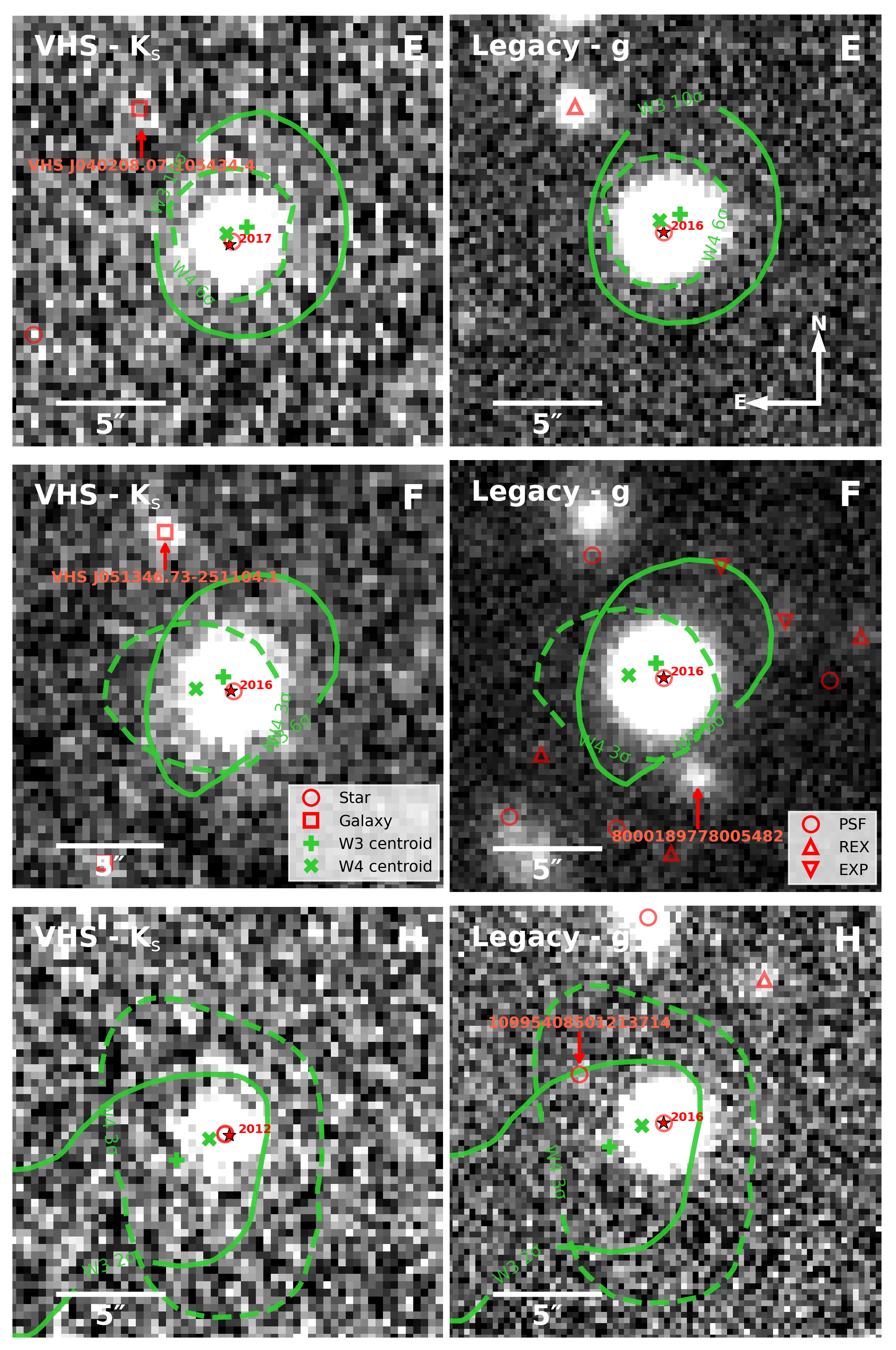}
    \caption{VHS $K_{\rm s}$-band (left panels) and Legacy $g$-band (right panels) images of candidates E, F and H. Red stars mark the propagated \emph{Gaia} positions at the VHS imaging epochs (2017 for E, 2016 for F, and 2012 for H), while yellow stars show the \emph{Gaia} positions at the \emph{Gaia} DR3 reference epoch (2016.0). Green plus signs ($+$) and crosses ($\times$) mark the $W3$ and $W4$ centroids, respectively. The corresponding green $W3$ (solid) and $W4$ (dashed) high flux contours for candidates are also shown, with all data measured in Section \ref{sec:section_3}. In the VHS panels, red circles show stellar objects from the VHS catalogue, and the red squares mark the galaxy-like objects. In the Legacy $g$-band panels, red circles indicate sources with PSF profiles, while red triangles mark REX sources using upward triangles and EXP sources using downward triangles. A compass in the upper-right panel (Legacy $g$-band) indicates the orientation for all subplots. Nearby companions are highlighted with red labels: VHS J040208.07$-$105434.4 in the $K_{\rm s}$ panel of candidate~E; VHS J051346.73$-$251104.1 ($K_{\rm s}$) and Legacy ID 8000189778005482 ($g$-band) for candidate~F; and Legacy ID 10995408501213714 in the $g$-band panel of candidate~H.}
    \label{fig:EFH_Images}
\end{figure}

As for the remaining candidates, E and F show conditions similar to candidate A (Figure~\ref{fig:EFH_Images}). They exhibit $W3$ and $W4$ centroids that are spatially coincident with the brightness profiles of the central stars. Additionally, no other objects appear to be located within the regions of MIR flux.

Candidate H presents a more complex morphology. While its $W4$ centroid aligns with the central stellar source, the $W3$ centroid is significantly offset by $\sim2.6\arcsec$, placing it outside the star's main brightness profile. Unlike candidate B, the multi-band images of candidate H do not reveal any extended structures that indicates a companion. However, inspection of the Legacy DR10 data reveals a faint source with a PSF-like profile (DESI Legacy ID 10995408501213714) located $\sim 5\arcsec$ to the northeast (Figure~\ref{fig:EFH_Images}). This object, likely a distant galaxy, coincides with the high-flux region of the $W4$ emission and lies on the fringe of the $W3$ emission.

\subsection{Image diagnostics for candidate C}
\label{sec:section_4_3}

\begin{figure}
    \centering
    \includegraphics[width=1.0\linewidth]{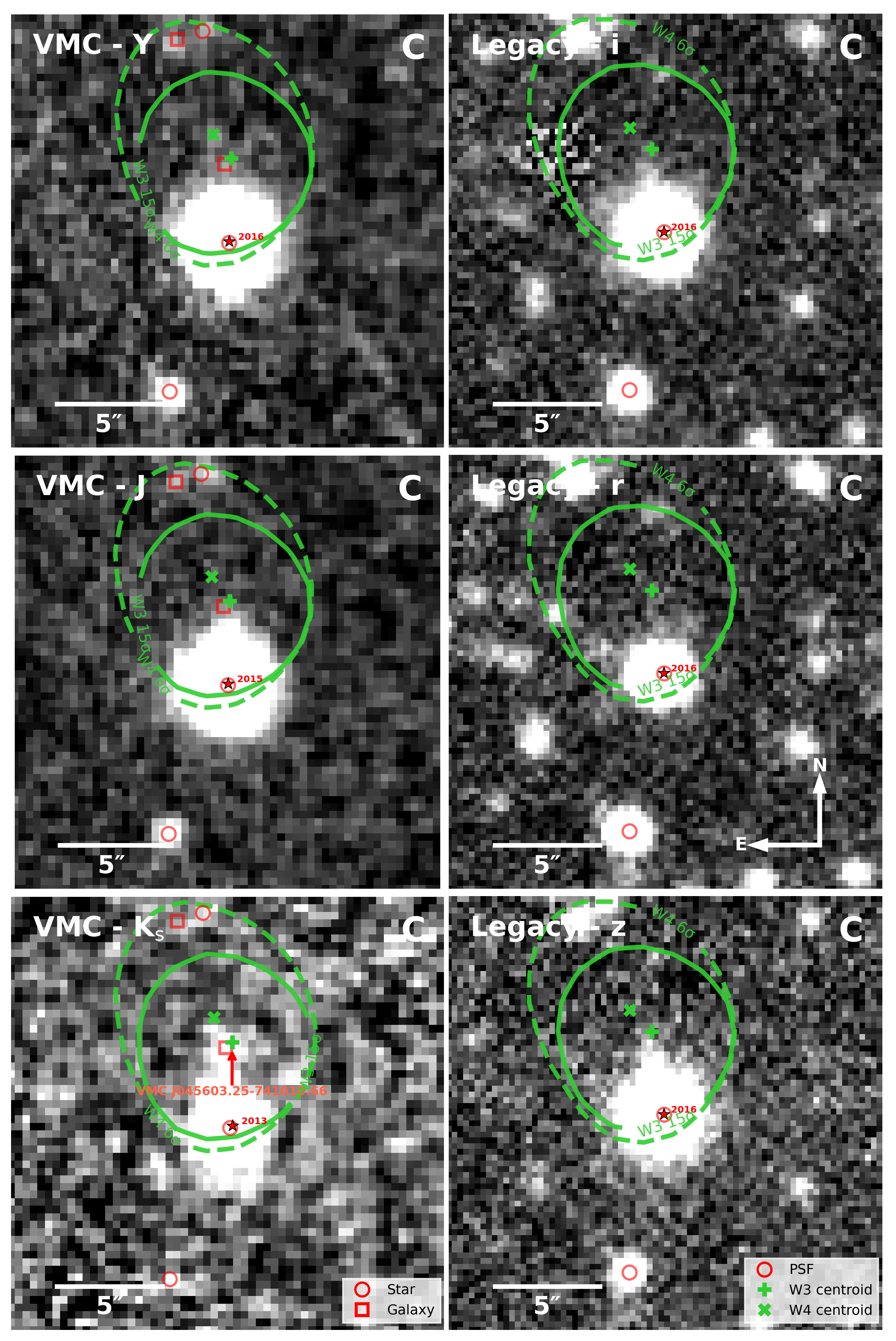}
    \caption{VMC $Y$-, $J$-, and $K_{\rm s}$-band (left panels) and Legacy DR10 $i$-, $r$-, and $z$-band (right panels) images of candidate C. Red stars mark the stellar position at the observation epochs of the VMC images, and the stellar positions at 2016.0 (the \emph{Gaia} DR3 epoch) for the Legacy images.  Green plus signs ($+$) and crosses ($\times$) mark the $W3$ and $W4$ centroids, respectively. The corresponding green $W3$ (solid) and $W4$ (dashed) high-flux contours for candidates are also shown, with all data measured in Section \ref{sec:section_3}. In the VMC panels, red circles denote stellar-like objects from the VMC catalogue, and the red squares mark the galaxy-like objects; in the Legacy panels, the red circles indicate sources best fit by PSF profiles. A compass in the middle-right panel (Legacy $r$-band) indicates the orientation for all subplots. VMC J045603.25-741010.66, highlighted by a red label, is visible north of the star, most significantly in the VMC $K_{\rm s}$ band.}
    \label{fig:C_VMC}
\end{figure}

\begin{figure}
    \centering
    \includegraphics[width=0.62\linewidth]{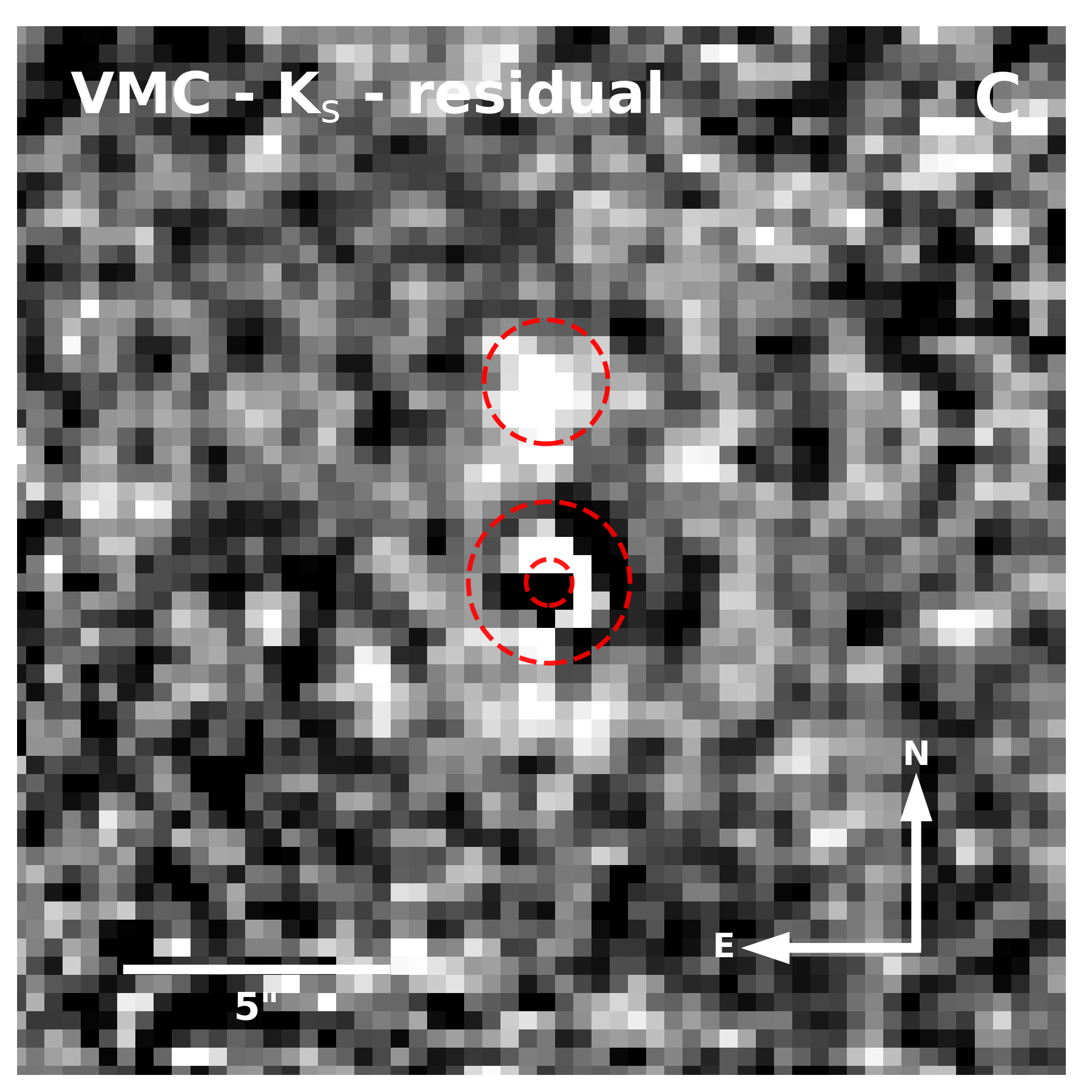}
    \caption{PSF residual images of candidate C in the VMC $K_{\rm s}$ band. The two central red dashed circles show the PSF FWHM and the 99\% brightness profile of the star. The offset red dashed circle marks VMC J045603.25-741010.66, the companion source detected outside the PSF brightness profile. North (N) and east (E) are indicated by the compass in the bottom-right corner.}
    \label{fig:C_residual}
\end{figure}

Candidate C is the only case covered by the VMC survey, and it lacks any associated radio or X-ray counterparts from archival data. Figure~\ref{fig:C_VMC} shows its VMC $J$-, $Y$-, and $K_{\rm s}$-band images, and the Legacy DR10 $r$-, $i$-, and $z$-band images. A critical finding is the presence of an extended tail-like structure located to the north of the central star, with a peak emission reaching a significance of $>5\sigma$. This tail corresponds to the catalogued source VMC J045603.25-741010.66, detected as a faint extended source with a $K_{\rm s}$-band AB magnitude of $21.60 \pm 0.32$~mag. The morphological analysis in the VISTA Magellanic Survey (VMC) catalogue ($YJK_{\rm s}$) DR6 \citep{2011A&A...527A.116C} classifies this source as a likely galaxy with a $90\%$ probability. We also validated this detection by inspecting the VMC $K_{\rm s}$-band PSF residual, which confirmed that the significant feature is distinct from the stellar profile (Figure~\ref{fig:C_residual}). Crucially, while VMC J045603.25-741010.66 is $3.50\arcsec$ away from the \emph{Gaia} position of candidate star C, it is only $0.4\arcsec$ from the WISE $W3$ centroid.

VMC J045603.25-741010.66, the companion source, lacks a significant counterpart in the VMC $J$- and $Y$-band images, as confirmed by both the VMC catalogue and our inspection of the corresponding image data. Furthermore, while the source remains faint in the Legacy DR10 $r$ and $i$ bands, its peak significance is detected at $\sim3\sigma$ above the background in the Legacy $z$-band image.

\subsection{Image diagnostics for candidates D, I, and J}
\label{sec:section_4_4}

\begin{figure}
    \centering
    \includegraphics[width=1.0\linewidth]{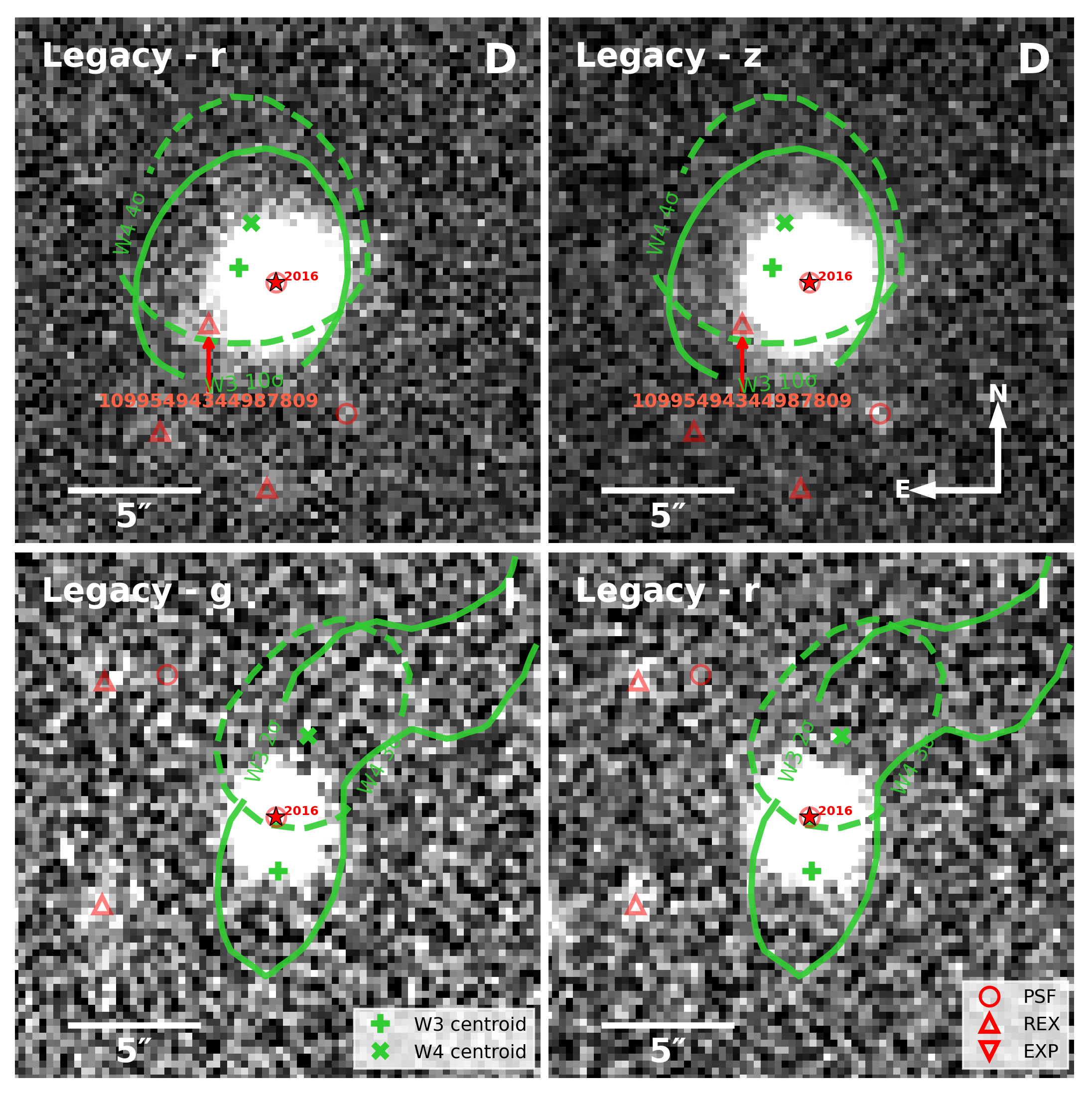}
    \caption{Legacy DR10 images for candidates D and I. The upper panels show the $r$-band and $z$-band images for candidate D, and the lower panels show the $g$-band and $r$-band images for candidate I. Each panel spans $\sim$20$\arcsec\times$20$\arcsec$ in RA and Dec. Red stars mark the \emph{Gaia} positions at the \emph{Gaia} DR3 epoch (2016.0). Green plus signs ($+$) and crosses ($\times$) mark the $W3$ and $W4$ centroids, respectively. The corresponding green $W3$ (solid) and $W4$ (dashed) high flux contours for the candidates are also shown, with all data measured in Section \ref{sec:section_3}. Red circles indicate sources with PSF profiles from the Legacy catalogue, red upward triangles mark REX sources, and red downward triangles mark EXP sources. North (N) and east (E) directions are indicated by the compass marker in the bottom-right corner of the upper-right panel and apply to all cutouts. One nearby companion, Legacy ID: 10995494344987809 for candidate D, is highlighted by annotation.}
    \label{fig:DI_legacy}
\end{figure}

\begin{figure*}
    \centering
    \includegraphics[width=1\linewidth]{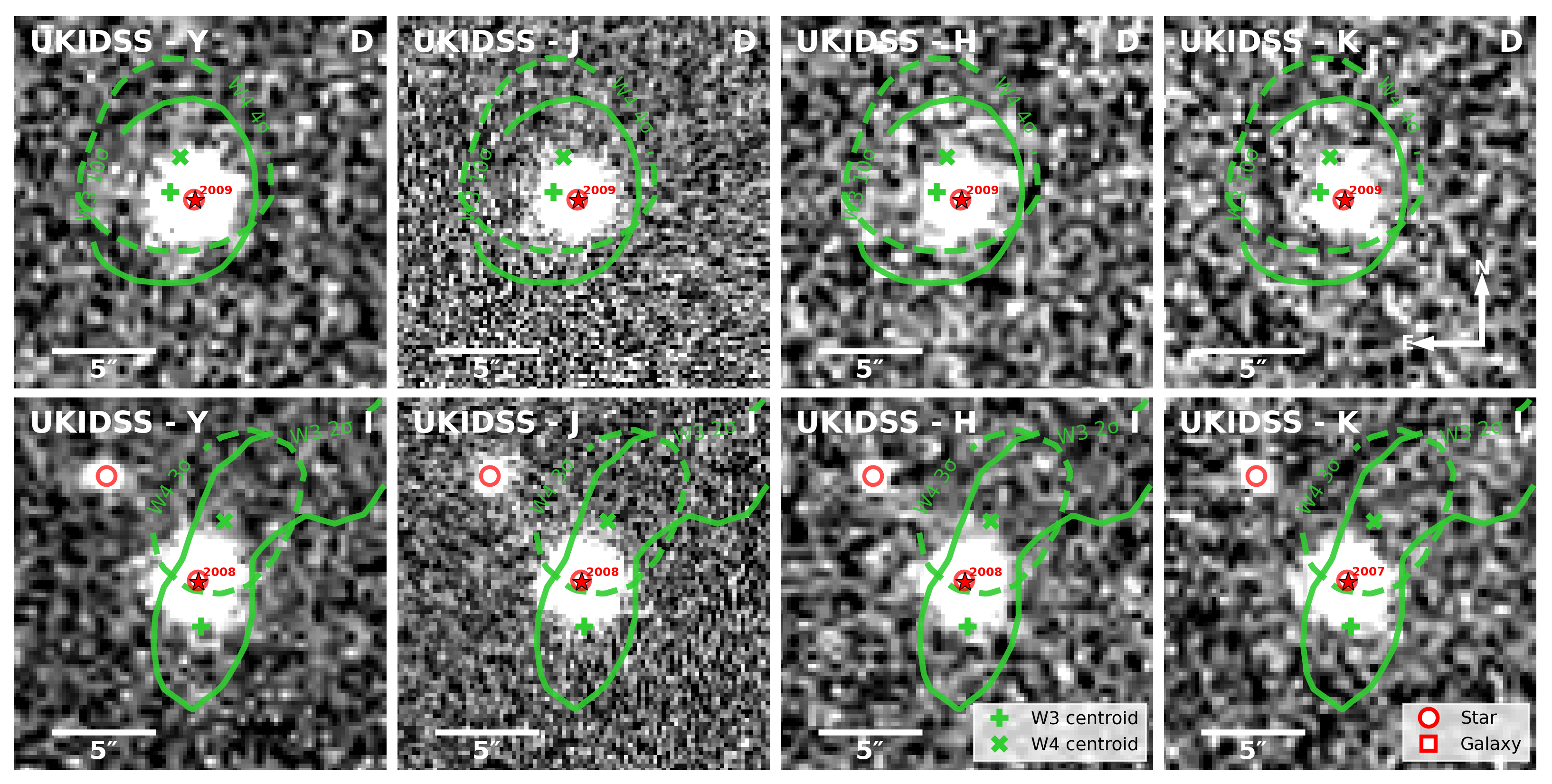}
    \caption{UKIDSS $Y$-, $J$-, $H$-, and $K$-band images for candidates D and I, with each subplot displaying an $18\arcsec\times18\arcsec$ field of view. The upper panels display data for candidate D observed between August and October 2009, while the lower panels display data for candidate I observed between April 2007 and February 2008. Red stars mark the propagated \emph{Gaia} positions at the precise observation epochs retrieved from the archive logs (annotated by year). Red circles indicate sources classified as stars within the companion survey catalogue. Green plus signs ($+$) and crosses ($\times$) mark the $W3$ and $W4$ emission centroids, respectively. The corresponding green solid lines outline the $W3$ flux contours, and the green dashed lines outline the $W4$ contours. North (N) and east (E) directions are indicated by the compass marker in the bottom-right corner of the upper-right panel and apply to all cutouts.}
    \label{fig:DI_UKIDSS}
\end{figure*}

Candidates D, I, and J have complete Legacy imaging coverage but lack VISTA data; instead, they are covered by the UKIRT surveys. In the NIR regime, we analyse D and I using UKIDSS imaging data, and J using the UHS. Figure~\ref{fig:DI_legacy} presents the Legacy images used to assess the optical morphology of candidates D and I, while the UKIDSS images for them are shown in Figure~\ref{fig:DI_UKIDSS}.

For candidate D, the Legacy catalogue reveals a faint companion object. With a $\text{REX}$ profile and DESI Legacy ID 10995494344987809, this object is located on the south-eastern fringe of the stellar brightness profile. This object appears as a marginal detection in the $r$-band image (Figure~\ref{fig:DI_legacy}), and is offset by $2.4\arcsec$ from the $W3$ centroid position. For candidate I, the $W3$ centroid detected at low SNR appears to lie on the fringe of the stellar brightness profile, while the $W4$ centroid lies outside the profile. Despite the positional separation suggested by the WISE images, the Legacy catalogue did not identify any clear companion object located within 3$\arcsec$ of the position of the candidate star, unlike the case of candidate D. In the UKIDSS images (Figure~\ref{fig:DI_UKIDSS}), there are even fewer companion objects or nearby stars in the vicinity of candidates D and I, and neither field reveals an obvious contaminating source.

\begin{figure*}
    \centering
    \includegraphics[width=1\linewidth]{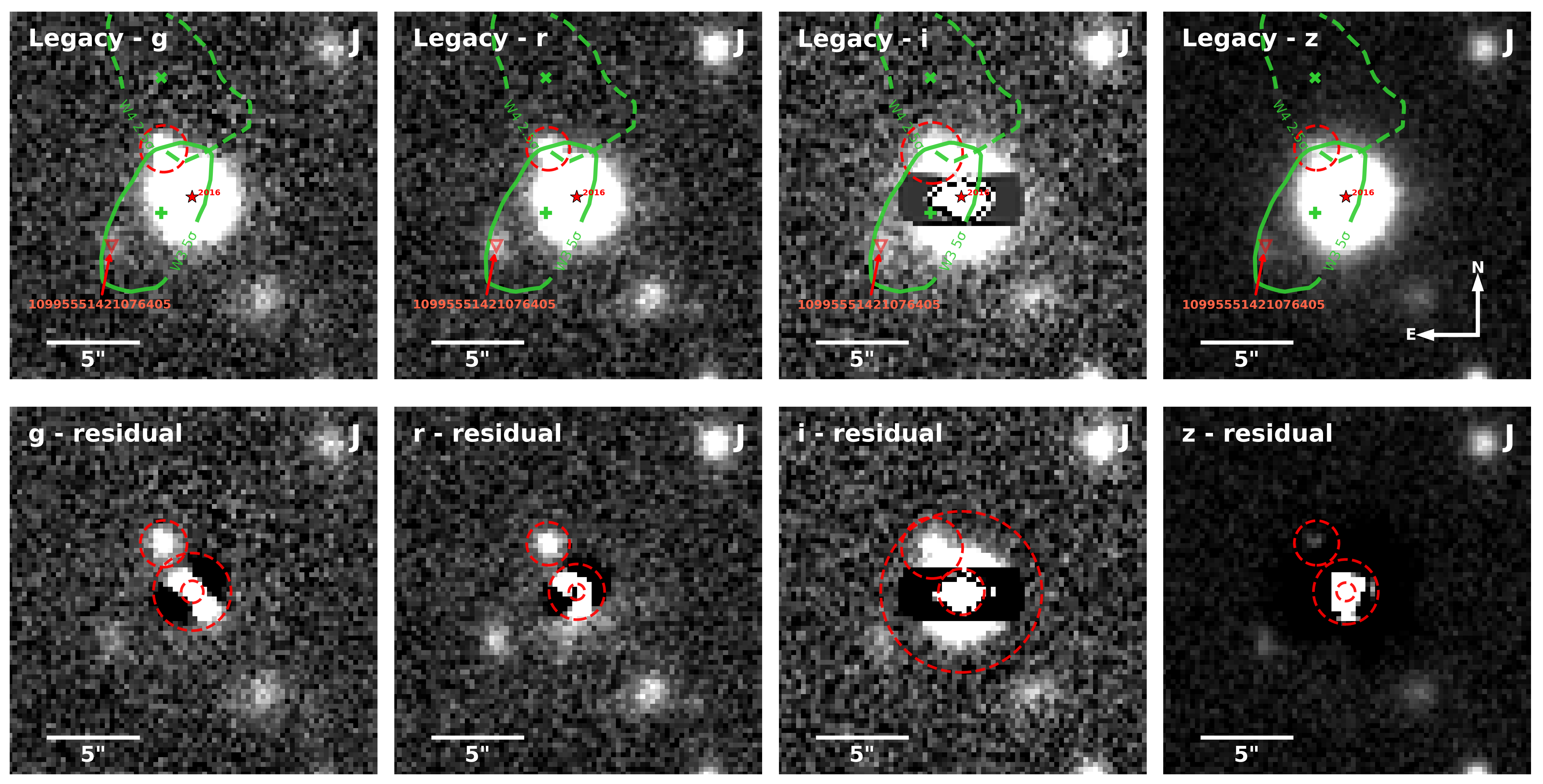}
    \caption{Legacy DR10 $g$-, $r$-, $i$-, and $z$-band images (upper panels) and corresponding PSF-subtracted residual images (lower panels) for candidate star~J, each displaying an $18\arcsec\times18\arcsec$ field of view. The upper panels show the original images, with standard WISE $W3$ emission centroids (green $+$) and $W4$ centroids (green $\times$) annotated. The solid green lines outline the $W3$ $5\sigma$ flux contours, and the dashed green lines outline the $W4$ $2.5\sigma$ contours. Red stars mark the \emph{Gaia} stellar positions at epoch J2016.0. A nearby catalogued source (Legacy ID:~10995551421076405) is additionally highlighted by the red marker, label, and arrow in the upper panels. North (N) and east (E) directions are indicated by the compass marker in the bottom-right corner of the upper-right panel and apply to all cutouts. The lower panels display the residual images following subtraction of a fitted circular PSF model of candidate~J. The two central red dashed circles show the PSF FWHM and the 99\% brightness profile of candidate~J, while the offset red dashed circle marks the position of the uncatalogued companion located $\sim 3\arcsec$ to the north-west. The companion is recovered as a significant residual in the $g$- and $r$-band images.}
    \label{fig:J_Legacy}
\end{figure*}

\begin{figure}
    \centering
    \includegraphics[width=1\linewidth]{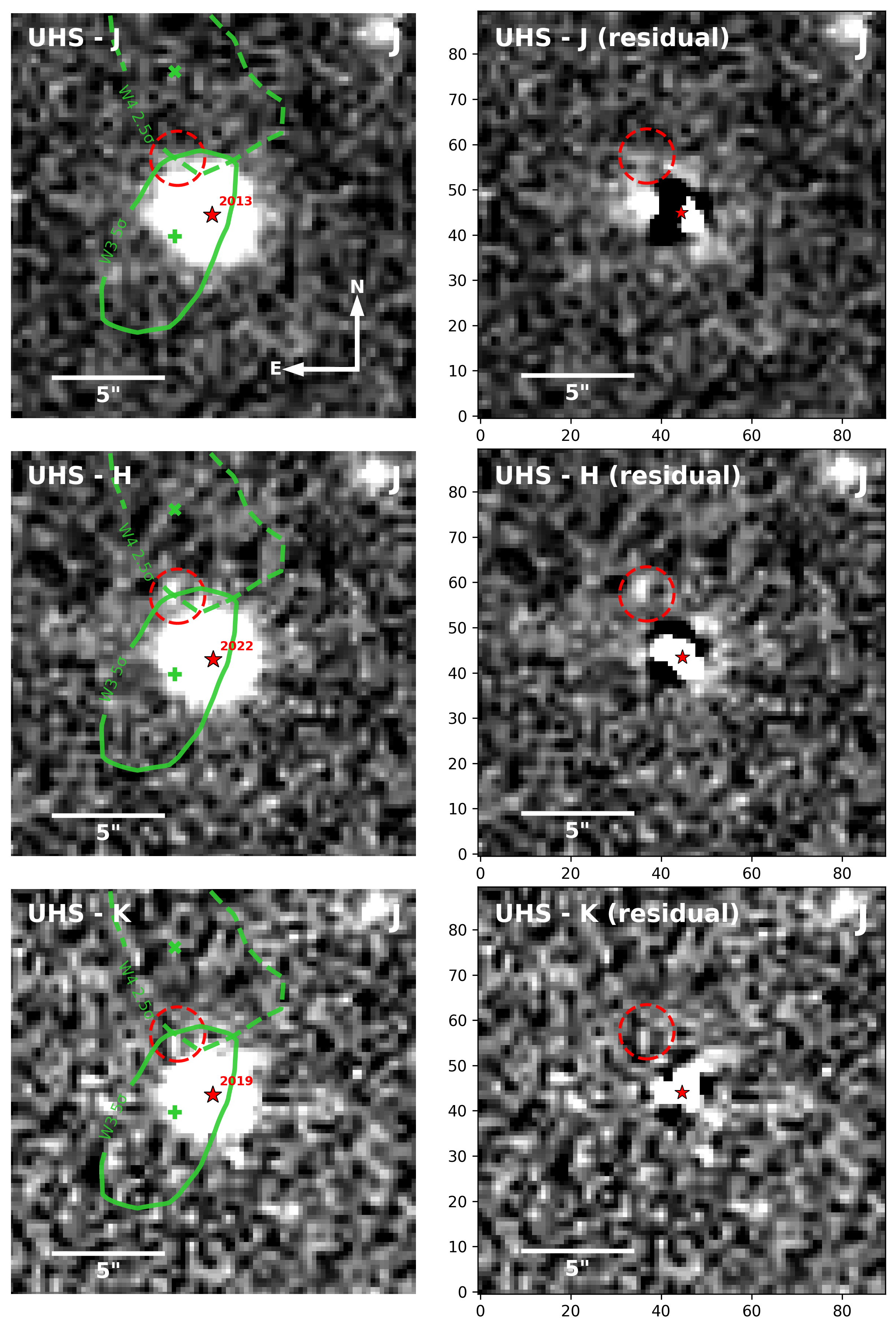}
    \caption{UHS DR2 NIR data and corresponding PSF-subtracted residual images for candidate star J across the $J$, $H$, and $K$ bands, each displaying an $18\arcsec\times18\arcsec$ field of view. The left panels show the original multi-band stacked images, with standard WISE $W3$ emission centroids (green $+$) and $W4$ centroids (green $\times$) mapped. The solid green lines outline the $W3$ $3\sigma$ flux contours, while the dashed green lines outline the $W4$ $2.5\sigma$ contours. Red stars mark the \emph{Gaia} positions propagated via proper motions to the per-band observation epoch (annotated by year: 2013, 2022, 2019). North (N) and east (E) directions are indicated by the compass marker in the bottom-right corner of the upper-left panel and apply to all cutouts. The right panels display the residual images following subtraction of a fitted circular PSF model, revealing persistent, non-stellar near-infrared emission coincident with the pre-computed $W4$ centroid location, as highlighted by the red dashed circle.}
    \label{fig:J_UHS}
\end{figure}

In contrast, candidate~J shows a blended morphology, with the Legacy image revealing a distinct bulge extending north-east from the stellar profile (Figure~\ref{fig:J_Legacy}). This feature also behaves as a faint companion, located at the same position in the UHS NIR band images, most notably in the $H$ band (Figure~\ref{fig:J_UHS}). However, this object is not catalogued as an independent source in either the Legacy or UHS catalogues. To evaluate its properties, we conducted photometric measurements across several optical and NIR bands. Figures~\ref{fig:J_Legacy} and \ref{fig:J_UHS} display the full-band Legacy and UHS maps, along with the residual maps of the vicinity after subtracting the stellar point spread function (PSF) of candidate J. Table~\ref{tab:J_companion} presents the measurement results.

\begin{table*}
\centering
\caption{Photometric extraction results of the optical/NIR companion source
located $3\arcsec$ from candidate~J, from Legacy DR10 ($g$, $r$, $i$, and $z$) and
UHS ($J$, $H$, and $K$) imaging.}
\label{tab:J_companion}
\begin{tabular}{c c c c c}
\hline
Band & AB magnitude & Flux density $F_\nu$ ($\mu$Jy) & Significance & Note \\
\hline
$g$ & $23.38 \pm 0.07$ & $1.61 \pm 0.11$ & $14.8\sigma$ &  \\
$r$ & $22.42 \pm 0.08$ & $3.92 \pm 0.29$ & $13.4\sigma$ & \\
$i$ & --- & --- & --- & excluded (saturation / bleed trail) \\
$z$ & $> 23.27$ & $< 11.27$ & $< 3\sigma$ & $\sim3.6\sigma$ residual peak \\
\hline
$J$ & $> 21.55$ & $< 8.71$  & $2.6\sigma$ & \\
$H$ & $21.05 \pm 0.28$ & $13.80 \pm 3.56$ & $3.9\sigma$ & tentative detection \\
$K$ & $> 20.74$ & $< 18.37$ & $0.3\sigma$ \\
\hline
\end{tabular}
\end{table*}

In the optical regime, this companion is most prominent in the $g$ and $r$ bands, where the photometric measurements are most reliable, yielding SNRs of 14.8 and 13.4, respectively (Figure~\ref{fig:J_Legacy}). The $i$-band image suffers from saturation and a bleed trail originating from candidate~J, permitting only an unreliable measurement, while the source is faint and undetected above $3\sigma$ in the $z$ band; these two bands are thus far less constraining than $g$ and $r$. We fitted the companion in the $g$- and $r$-band residuals and found that a PSF model best describes its morphology, meaning the companion is consistent with a point source and cannot be spatially resolved at this resolution.

We also performed photometric extraction in the UHS NIR bands (Figure~\ref{fig:J_UHS}). The $H$ band provides the best detection, recovering a $3.9\sigma$ source with an elongated, bar-like residual structure. In the $J$ band, a nominal $2.6\sigma$ peak is present but is indistinct from the fringe features of the candidate star's brightness profile. In the $K$ band, an elongated structure is again visible, but its significance is too low (${\rm SNR} = 0.3$) to make a convincing detection. Neither the $J$ nor the $K$ band provides a clear detection comparable to that in the $H$ band or to those in the $g$ and $r$ bands from the Legacy imaging.

This bulge-like companion lies $\sim 3\arcsec$ from both the $W3$ and $W4$ centroids. The Legacy DR10 also identified an $\text{EXP}$ profile source (DESI Legacy ID 10995551421076405) $\sim5\arcsec$ from candidate~J to the south-east of the $W3$ centroid. The complexity of the vicinity of candidate~J, together with the results for the remaining candidates, is discussed and synthesised in Section~\ref{sec:section_5}.

\section{Discussion}
\label{sec:section_5}

\subsection{Evaluating the Likelihood of Contamination for the Nine Candidates}
\label{sec:section_5_1}

Building on the multi-wavelength and astrometric diagnostics presented in Sections \ref{sec:section_3} and \ref{sec:section_4}, we can now classify the nine Dyson Sphere star candidates based on the likelihood of extragalactic contamination. Similar to the analysis of candidate G, we integrate these archival findings to identify which sources show clear signs of background contaminants and which remain ambiguous.

\subsubsection{Candidates B and C: the most likely contaminated cases}

Candidates B and C represent the two cases exhibiting the strongest evidence of background contaminants contributing to the MIR excess. For both cases, archival data reveal the presence of background objects consistent with the observed infrared anomalies. For candidate~B, additional astrometric evidence comes from the $W2$ band: compared to control-sample M dwarfs of similar $G$-band magnitude and $W2$ SNR, its emission centroid shows a larger offset from the propagated \emph{Gaia} position, with a Gaussian-equivalent significance of $>2\sigma$ under both scales (Table~\ref{tab:rayleigh}).

As shown in Figures~\ref{fig:AB_WISE} and \ref{fig:AB_Images}, the 1367.5~MHz and 1655.5~MHz radio sources of candidate B, along with the $W3$ and $W4$ centroids, are all spatially associated with the faint, extended tail/bulge structure visible in the Legacy Survey images, lying at the fringe of candidate B's PSF brightness profile. The spatial alignment, reinforced by the typical AGN radio spectral index of $\alpha \sim 0.63$ presented in Subsection~\ref{sec:section_3_1}, supports the presence of a background extragalactic source blended with the foreground candidate B. The WISE emission maps (Figure~\ref{fig:AB_WISE}) suggest that this background galaxy is likely a hot DOG, which dominates the observed MIR flux in the $W3$ and $W4$ bands. At WISE's angular resolution, the $3\arcsec$ galaxy-to-stellar offset creates the illusion that candidate B itself is emitting the $W3$ and $W4$ excess fluxes, consistent with the predicted MIR signature of a modelled Dyson Sphere star.

Although candidate C lacks radio counterparts, its $W3$ and $W4$ emissions are offset from the central star by several arcseconds, similar to candidates B and G. The $W3$ and $W4$ centroids are spatially coincident with VMC J045603.25-741010.66, a source detected independently in the VMC and Legacy surveys. Given that its location is nearly identical to the $W3$ centroid measured in our work, and there is no other nearby object detected, VMC J045603.25-741010.66 is likely the genuine contributor to the MIR flux excesses in the $W3$ and $W4$ bands of candidate C. The WISE photometric characteristics of VMC J045603.25-741010.66 suggest that it is a potential hot DOG previously unknown to the literature. However, it is undetected in both the RACS and GLEAM radio surveys. This non-detection is consistent with the typically radio-quiet nature of the hot DOG population \citep{2013ApJ...769...91B,2016MNRAS.455.2058F}. We further discuss the implications of this result and evaluate the expected radio-loud fraction of hot DOGs based on a sample of 144 confirmed hot DOGs in Section \ref{sec:section_5_2}.

\subsubsection{Candidate A: a suggestive contaminated case}

Candidate A represents a highly suggestive case of contamination by a background AGN. Unlike candidates B and C, which have catalogued companions spatially separated from the position of the known star, candidate A’s radio companion (RACS-DR1 J124512.7-265206) is blended with the star position and thus not distinct. The radio spectral index of RACS-DR1 J124512.7-265206 ($\alpha \sim 0.40$) is characteristic of a flat-spectrum AGN. The radio sources, WISE centroids, and the \emph{Gaia} positions of the candidate star A are highly coincident, with mutual offsets of less than $2.5\arcsec$. As shown in Figure~\ref{fig:AB_Images}, these positions align closely within the brightness profile of the candidate star, and archival optical/NIR imaging reveals no other nearby sources. Given the statistical rarity of a flat-spectrum radio source aligning so precisely with a foreground star, it is highly probable that the star's emission is being contaminated by a background AGN along the same line of sight. This alignment explains the minimal positional offsets observed across the different wavebands. Direct confirmation, however, requires high-resolution observations. Upcoming high-resolution observations from the \emph{JWST} GO programme 7199 \citep{2025jwst.prop.7199Z} using the MIRI instrument will be essential to disentangle the mid-infrared emission from the stellar profile.

\subsubsection{Candidates H and J: the candidates with ambiguous evidence}

Unlike candidates~A, B, and~C, for which potential contamination sources were clearly identified, candidate~H represents an ambiguous case. Its $W3$-band centroid is offset by $\sim2.6\arcsec$ from the \emph{Gaia} position of the star, the third-largest $W3$-band offset among the candidates after candidates~B and C. However, its $W1$- and $W2$-band offsets are only $\sim0.5\arcsec$. Although the quantitative Rayleigh test in Subsection~\ref{sec:section_3_5} suggests that the $W3$ offset is not significantly different from the control star distribution, the control sample contains only five stars. Therefore, this statistical result should not obscure the reality of the high positional $W3$-band offset of candidate H, which is clearly visible in Figure~\ref{fig:HIJ_WISE}. Although its $W4$ centroid offset is smaller ($<2.5\arcsec$), the spatial MIR emission morphology appears elongated.

Using Legacy Survey data, we identified a faint object with a PSF brightness profile---likely a background galaxy (Legacy ID: 10995408501213714)---situated to the northeast of candidate star H that contributes to the $W4$ flux (Figure~\ref{fig:EFH_Images}). The blending of these two sources may result in the observed elongation of the $W4$ emission region. However, the SNRs of the $W3$ and $W4$ emission from candidate H are low ($\sim3\sigma$), significantly undermining the reliability of its measured offsets. Furthermore, unlike candidate C, where the catalogued background galaxy is nearly coincident with the $W3$-band centroid, the catalogued background galaxy near candidate H lies $\sim3\arcsec$ from the $W3$ centroid.

Another ambiguous case is candidate~J. As detailed in Section~\ref{sec:section_4_4}, we retrieved photometric evidence for an uncatalogued companion source at the stellar fringe, detected consistently in the Legacy $g$ and $r$ bands and tentatively in the UHS $H$ band, and exhibiting a point-source morphology. This companion is not, however, the only source projected onto the $W3$ emission of candidate~J: it lies $3.35\arcsec$ from the $W3$ centroid and $3.94\arcsec$ from the $W4$ centroid, while the Legacy DR10 Tractor catalogue separately lists a resolved, galaxy-like source with an exponential (EXP) profile (Legacy ID 10995551421076405; Figure~\ref{fig:J_Legacy}). As with candidate~H, both putative contributors to the MIR flux are offset from the $W3$ centroid by $\sim3\arcsec$ rather than being coincident with it.

Since the candidate star~J, the companion, and the catalogued galaxy-like source (Legacy ID: 10995551421076405) all lie within the $W3$ $>5\sigma$ emission region, the $W3$ flux cannot be confidently attributed to a single source: owing to the blending and co-location within the WISE beam, it may plausibly be dominated by the star, by the companion, by the galaxy-like source, or by a combination of these objects. The $W4$ emission, by contrast, exhibits a significant positional offset from both the candidate star and the catalogued galaxy-like source, leaving the companion as the only optical/NIR feature that could be preliminarily identified as a contributor to the $W4$ flux. We accordingly regard candidate~J as an ambiguous case of background contamination; unlike candidates~A and B, whose blended companion can be identified as an AGN through its radio spectral index, the companion of candidate~J lacks comparably direct evidence for its nature, and definitively assigning the MIR emission to any single feature will require future observations of higher angular resolution and sensitivity. 

Given the lack of exact spatial coincidence between the newly identified companion sources and the $W3$ centroids, the evidence for candidates~H and~J is substantially less compelling than that for candidates~A, B, and~C. Neither candidate shows a centroid-to-stellar offset that is significant relative to its control sample (Table~\ref{tab:rayleigh}), so their classification rests on the presence of a plausible contaminant rather than on a confirmed match to the MIR centroid. We therefore place candidates~H and~J in a lower, ambiguous classification tier.

\subsubsection{Candidates that lack critical evidence for MIR contamination}

Candidates D, E, F, and I lack both clear evidence of contamination and the definitive proof needed for a final classification. Without radio detections, their analysis relies solely on ALLWISE centroids and archival optical/NIR imaging analysis, following the same methodology used for candidates C, H, and J.

Candidate D presents a spatial contradiction: while Legacy imaging reveals an \text{REX}-profile galaxy (Legacy ID: 10995494344987809) on the southeastern fringe of the stellar profile (Figure~\ref{fig:DI_legacy}), the WISE centroids are oriented toward the north. Although its $W3$ and $W4$ centroids lie $\sim 1.3\arcsec$ and $\sim2.3\arcsec$ from the star (Table~\ref{tab:CDEF_WISE}), no band shows an offset that is significant relative to the control population under either the Rayleigh or the median scale (Table~\ref{tab:rayleigh}). A plausible explanation for the low-level excess is a proximate, highly blended AGN---analogous to the blending of candidate~A, but without radio emission to confirm an extragalactic nature. Despite candidate D lying within the footprint of multiple major radio surveys ($\delta \approx +05^\circ$), the absence of a radio counterpart suggests that any contaminant would have to be radio-quiet. 

Candidates~E and~F are weaker still. Like candidate~D, neither passes the Rayleigh or median test in any band (Table~\ref{tab:rayleigh}), and their $W3$ and $W4$ offsets are smaller than D's ($1.15\arcsec$ and $0.81\arcsec$ for E; $0.70\arcsec$ and $1.69\arcsec$ for F; Table~\ref{tab:CDEF_WISE}). Deep imaging reveals no plausible contaminant close enough to affect the MIR photometry: the nearest catalogued sources are VHS~J040208.07$-$105434.4 for candidate~E and VHS~J051346.73$-$251104.1 for candidate~F, both more than $5\arcsec$ away (Figure~\ref{fig:EFH_Images}).

The final case is candidate I. Its $W3$ centroid appears to be located on the fringe of the stellar brightness profile, while the $W4$ centroid is positioned outside the profile. The $W3$ emission, however, displays a stripe-like structure with only $\sim 2\sigma$ significance, making it an unreliable indicator of the true background-source position. Its $W4$ emission, with $>3\sigma$ significance, shows an offset of $\sim3.2\arcsec$ from the \emph{Gaia} position of the star. However, the comparison in Subsection~\ref{sec:section_3_5} does not support a significant offset relative to the control stars in any band; the only marginal value, a median-scale significance of $2.18\pm3.56\sigma$ in $W2$, is too poorly constrained to be meaningful. Unlike candidates~H and~J, for which deep optical/NIR imaging revealed sources projected onto the high $W3$ and/or $W4$ flux regions, we find no such counterpart for the MIR emission of candidate~I in the deep Legacy Survey and UHS images. We therefore place candidate~I in the lowest classification tier, alongside candidates~D, E, and~F.

Ultimately, the true origin of the emission from candidates~D and~E will be determined by high-resolution MIRI data from the \emph{JWST} GO~7199 programme \citep{2025jwst.prop.7199Z}, whereas candidates~F and~I are not included in this programme. The potential contaminants and classifications for all candidates discussed in Subsection~\ref{sec:section_5_1} are summarised in Table~\ref{tab:hot_dog_contaminants}.

\newcommand{\cmark}{\checkmark}
\newcommand{\xmark}{$\times$}

\begin{table*}
    \centering
    \caption{Summary of potential background contaminants among the Dyson Sphere candidates.}
    \label{tab:hot_dog_contaminants}
    \begin{tabular}{c c c c}
        \hline
        Candidate & Evidence Strength & Likely Contaminant ID & Note \\
        \hline
        A & Suggestive & RACS-DR1 J124512.7-265206 & Exhibits AGN-like radio spectrum \\
        B & Strong     & RACS-DR1 J035603.8-403148 & Exhibits AGN-like radio spectrum \\
        C & Strong     & VMC J045603.25-741010.66  &  \\
        D & Weak       & --                        &  \\
        E & Weak  & --                        &  \\
        F & Weak  & --                        &  \\
        H & Ambiguous  & Legacy 10995408501213714  &  \\
        I & Weak       & --                        &  \\
        J & Ambiguous & Legacy 10995551421076405  &  An uncatalogued companion source identified \\
        \hline
    \end{tabular}
\end{table*}

\subsection{The Radio Detection Rate of hot Dust-Obscured Galaxies}
\label{sec:section_5_2}

Among the ten Dyson Sphere candidates analysed in this work, only three (A, B, and G) were detected in archival radio surveys. If one assumes that all ten candidates are contaminated by hot DOGs, the remaining seven would necessarily be radio-quiet AGNs. To evaluate the plausibility of this scenario, we examined a sample of 144 catalogued hot DOGs from \citet{2015ApJ...804...27A}—a robust sample selected from WISE data and verified with ground-based NIR observations.

We queried the VizieR service for radio counterparts within this reference sample and identified only 12 detections (Table~\ref{tab:radio_detections}), representing a detection rate of $<10\%$. The radio catalogues include: the FIRST survey (\citealt*{1995ApJ...450..559B}), NVSS (\citealt{1998AJ....115.1693C}), SUMSS (\citealt{2003MNRAS.342.1117M}), the TIFR GMRT Sky Survey (TGSS; \citealt{2017A&A...598A..78I}), the Australia Telescope Compact Array XXL survey (ATCA XXL; \citealt{2018A&A...620A..16B}), the VLA Sky Survey (VLASS; \citealt{2020PASP..132c5001L}), the RACS surveys (\citealt{2021PASA...38...58H, 2024PASA...41....3D, 2025PASA...42...38D}), the LOFAR Two-metre Sky Survey Data Release 2 (LoTSS DR2; \citealt{2022A&A...659A...1S}), and the GLEAM-X survey (\citealt{2024PASA...41...54R}). This low detection rate is consistent with the characterisation of similar WISE-selected dusty extragalactic sources as typically radio-quiet \citep{2013ApJ...769...91B}. Consequently, the lack of radio emission in seven out of ten candidates aligns with the low radio-loud fraction expected for this galaxy class.

\begin{table}
\centering
\caption{Radio detections for the sample of 144 catalogued hot DOGs.}
\label{tab:radio_detections}
\scriptsize
\begin{tabular}{cccc}
\hline
WISE ID & Survey & Central Frequency & Flux Density (mJy) \\
\hline
J022052.12+013711.6
& VLA$^{a}$ & 6.2\,GHz & $0.20 \pm 0.01$ \\
\hline
\multirow[c]{3}{*}{J022646.85+051422.7}
& NVSS  & 1.4\,GHz & $6.10 \pm 0.50$ \\
& FIRST & 1.4\,GHz & $6.49 \pm 0.12$ \\
& VLASS & 3.0\,GHz & $3.22 \pm 0.23$ \\
\hline
\multirow[c]{2}{*}{J075725.07+511319.7}
& FIRST & 1.4\,GHz & $3.57 \pm 0.15$ \\
& NVSS  & 1.4\,GHz & $3.20 \pm 0.50$ \\
\hline
J085124.78+314856.1
& LoTSS DR2 & 144\,MHz & $5.22 \pm 2.83$ \\
\hline
\multirow[c]{2}{*}{J094608.30+500820.4}
& LoTSS DR2 & 144\,MHz & $6.64 \pm 1.16$ \\
& FIRST     & 1.4\,GHz & $1.21 \pm 0.16$ \\
\hline
\multirow[c]{5}{*}{J100636.34+510500.7}
& LoTSS DR2 & 144\,MHz & $31.95 \pm 9.64$ \\
& TGSS      & 147\,MHz & $40.0 \pm 9.0$ \\
& FIRST     & 1.4\,GHz & $4.13 \pm 0.13$ \\
& NVSS      & 1.4\,GHz & $3.20 \pm 0.40$ \\
& VLASS     & 3.0\,GHz & $2.02 \pm 0.25$ \\
\hline
\multirow[c]{4}{*}{J114612.87+412914.4}
& LoTSS DR2 & 144\,MHz & $4.38 \pm 1.89$ \\
& FIRST     & 1.4\,GHz & $4.39 \pm 0.13$ \\
& NVSS      & 1.4\,GHz & $4.10 \pm 0.40$ \\
& VLASS     & 3.0\,GHz & $2.39 \pm 0.20$ \\
\hline
J145705.92+293231.6
& LoTSS DR2 & 144\,MHz & $2.10 \pm 1.67$ \\
\hline
\multirow[c]{3}{*}{J154840.15+263029.3}
& RACS-low & 850\,MHz    & $2.27 \pm 0.46$ \\
& RACS-mid & 1367.5\,MHz & $1.97 \pm 0.41$ \\
& FIRST    & 1.4\,GHz    & $0.96 \pm 0.14$ \\
\hline
\multirow[c]{6}{*}{J160357.39+274553.2}
& RACS-low  & 850\,MHz    & $3.75 \pm 0.66$ \\
& RACS-mid  & 1367.5\,MHz & $2.10 \pm 0.50$ \\
& FIRST     & 1.4\,GHz    & $2.28 \pm 0.14$ \\
& NVSS      & 1.4\,GHz    & $2.80 \pm 0.50$ \\
& RACS-high & 1655.5\,MHz & $1.68 \pm 0.39$ \\
& VLASS     & 3.0\,GHz    & $0.86 \pm 0.17$ \\
\hline
\multirow[c]{3}{*}{J171400.14+524231.6}
& FIRST & 1.4\,GHz & $2.76 \pm 0.13$ \\
& NVSS  & 1.4\,GHz & $2.80 \pm 0.40$ \\
& VLASS & 3.0\,GHz & $1.30 \pm 0.25$ \\
\hline
\multirow[c]{6}{*}{J233953.74-561301.1}
& GLEAM     & 200\,MHz    & $86 \pm 0.50$ \\
& SUMSS     & 843\,MHz    & $21.2 \pm 1.1$ \\
& RACS-low  & 850\,MHz    & $17.84 \pm 0.40$ \\
& RACS-mid  & 1367.5\,MHz & $11.28 \pm 0.81$ \\
& RACS-high & 1655.5\,MHz & $9.05 \pm 0.97$ \\
& ATCA XXL  & 2.1\,GHz    & $7.70 \pm 0.39$ \\
\hline
\end{tabular}
\begin{tablenotes}
\small
\item $^{a}$\citet{2018MNRAS.477..830H}.
\end{tablenotes}
\end{table}

\section{Conclusions}
\label{sec:section_6}

We conducted a combined analysis of nine \emph{Project Hephaistos} Dyson Sphere candidates using two complementary approaches based on archival data: (1) an astrometric examination of AllWISE four-band image centroids to identify mid-infrared (MIR) flux offsets relative to the \emph{Gaia} positions propagated to the ALLWISE epoch, a method whose effectiveness can be validated through a comparative analysis of ordinary M-dwarf control samples, and (2) a search for potential extragalactic companions in deep NIR and optical images.

By comparing the WISE centroids with the \emph{Gaia} stellar positions (the first method), we found that candidates B and C exhibit the largest positional offsets, with $W3$ centroid-to-stellar offsets of approximately $3\arcsec$. Notably, the $W3$ centroid of candidate B aligns with RACS 1367.5~MHz and 1655.5~MHz radio counterparts to within $<1\arcsec$, indicating that a radio source dominates the emission observed in the $W3$ and $W4$ bands.

Using the second method---searching for extragalactic companions in deep imaging, we confirmed that J2335-0004, the hot DOG that contaminated candidate G, is present and catalogued in the VISTA and Legacy surveys. Although no radio companion was found near candidate C, we found a faint NIR source located at the position of the $W3$ centroid in the VMC data. These results suggest that the contaminants toward both B and C are likely hot DOGs, similar to the MIR positional offset observed in candidate G, which was later confirmed as a hot DOG by high-resolution radio observations \citep*{2025MNRAS.538L..56R}.

We also identified candidate~A as a suggestive case: it has a radio counterpart with an AGN-like spectral index, and the close alignment of the \emph{Gaia}, radio, and WISE positions implies that any contaminating background source must lie almost directly along the line of sight.

Candidates H and J are placed in a lower-confidence, ambiguous class requiring further evidence. Candidate~H shows one of the largest $W3$-band offsets in the sample, comparable to that of candidate~B, although it is not significant relative to its control stars; candidate~J has multiple possible background contaminants nearby. For both, plausible contaminating sources were identified, but they are not spatially coincident with the MIR centroids.

Lastly, candidates D, E, F, and I show no radio counterparts, no unambiguously outstanding centroid offsets, and no faint companions that could plausibly contribute to the MIR excess. However, we suspect that these sources might also be contaminated by radio-quiet AGNs that lie close to the line of sight towards the M-dwarf star. Our identification of a sample of 144 hot DOGs suggests that more than 90\% of them are radio-quiet. This result agrees with the low radio-detection rate observed in the \emph{Project Hephaistos} Dyson Sphere candidate sample.

Despite the resolution limitations of current MIR datasets like the WISE archive, our analysis demonstrates that supplementary multi-wavelength data offer valuable insights into the nature of these Dyson Sphere candidates. The suspected contaminants near candidates B and C still require confirmation of their extragalactic nature. Dedicated observations, such as deep infrared imaging and spectral analysis, are essential to conclusively disentangle hypothetical hot DOGs from the foreground stars. Such observations are equally critical for resolving the possible cases (A, H, and J) and clarifying the nature of candidates D, E, F, and I, where radio data remain insufficient. In this regard, the ongoing \emph{JWST} GO 7199 observations of candidates A, D, and E will provide the direct results needed to identify the genuine contributors to their infrared excess. Ultimately, rigorously identifying and eliminating such false positives is a fundamental step in the search for techno-signatures; by systematically ruling out natural contamination, we can ensure that future resources focus only on the most robust and compelling Dyson Sphere candidates.

\section*{Acknowledgements}

We sincerely thank the referee for the helpful suggestions, which
significantly improved the manuscript. TR and MAG applied the centroid analysis originally developed for their investigation of candidate G. EZ first noted the faint, diffuse feature near candidate C, motivating the development of the deep-image inspection method used in this
study. AB was supported by the Swedish National Space Agency. TR is funded by the China Scholarship Council.

This research made use of data and services provided by AllWISE \citep{2014yCat.2328....0C, 2020ipac.data.I153W}, the Australia Telescope Compact Array XXL survey (ATCA XXL; \citealt{2018A&A...620A..16B}), the Dark Energy Survey (DES; \citealp{2016MNRAS.460.1270D}; DES DR1: \citealp{2018ApJS..239...18A}; DES DR2: \citealp{2021ApJS..255...20A}), the DESI Legacy Imaging Surveys \citep{2019AJ....157..168D}, the FIRST survey (\citealt*{1995ApJ...450..559B}), {\it Gaia} \citep{2016A&A...595A...1G, 2023A&A...674A...1G}, the GLEAM-X survey (\citealt{2024PASA...41...54R}), the LOFAR Two-metre Sky Survey Data Release 2 (LoTSS DR2; \citealt{2022A&A...659A...1S}), NVSS (\citealt{1998AJ....115.1693C}), the Panoramic Survey Telescope and Rapid Response System (Pan-STARRS; \citealp{2016arXiv161205560C}), the RACS surveys (\citealt{2021PASA...38...58H, 2024PASA...41....3D, 2025PASA...42...38D}), SkyMapper \citep{2024PASA...41...61O}, SUMSS (\citealt{2003MNRAS.342.1117M}), the TIFR GMRT Sky Survey (TGSS; \citealt{2017A&A...598A..78I}), the Two Micron All Sky Survey (2MASS) \citep{2006AJ....131.1163S}, the Very Large Array Sky Survey (VLASS; \citealt{2020PASP..132c5001L}), and the VISTA Science Archive \citep{2012A&A...548A.119C}.

This research also made use of the Aladin Sky Atlas \citep{2000A&AS..143...33B}, \textsc{Simbad} \citep{2000A&AS..143....9W}, CASA\citep{2022PASP..134k4501C}, CARTA\citep{2026PASP..138b4506W}, ESASky \citep{2017PASP..129b8001B, 2018A&C....24...97G}, SAOImage DS9\citep{2003ASPC..295..489J}, the VizieR catalogue access tool, the NASA High Energy Astrophysics Science Archive Research Center, \textsc{Astroquery} \citep{2019AJ....157...98G},
\textsc{Astropy}\citep{astropy:2013, astropy:2018, astropy:2022}, \textsc{NumPy} \citep{harris2020array}, Photutils \citep{larry_bradley_2024_12585239}, \textsc{Matplotlib} \citep{2007CSE.....9...90H}, SEP \citep{barbary2016sep}, SciPy \citep{2020SciPy-NMeth}.

\section*{Data Availability}

All observational data analysed in this study are publicly available from the archives and survey data releases cited in the manuscript, including the NASA/IPAC Infrared Science Archive, the DESI Legacy Imaging Surveys, the VISTA Science Archive, the WFCAM Science Archive, and the corresponding public radio-survey data services. The catalogue identifiers, centroid positions, positional offsets, photometric measurements, and other derived quantities used in the analysis are reported in this article and its tables. The software used in the analysis is publicly available and cited in the manuscript. Additional intermediate products generated for this study will be made available by the corresponding author upon reasonable request.



\bibliographystyle{mnras}
\bibliography{example} 




\appendix

\section{Centroid-offset distributions of the control samples}
\label{sec:appendix}

\begin{figure*}
\centering
\includegraphics[width=\textwidth]{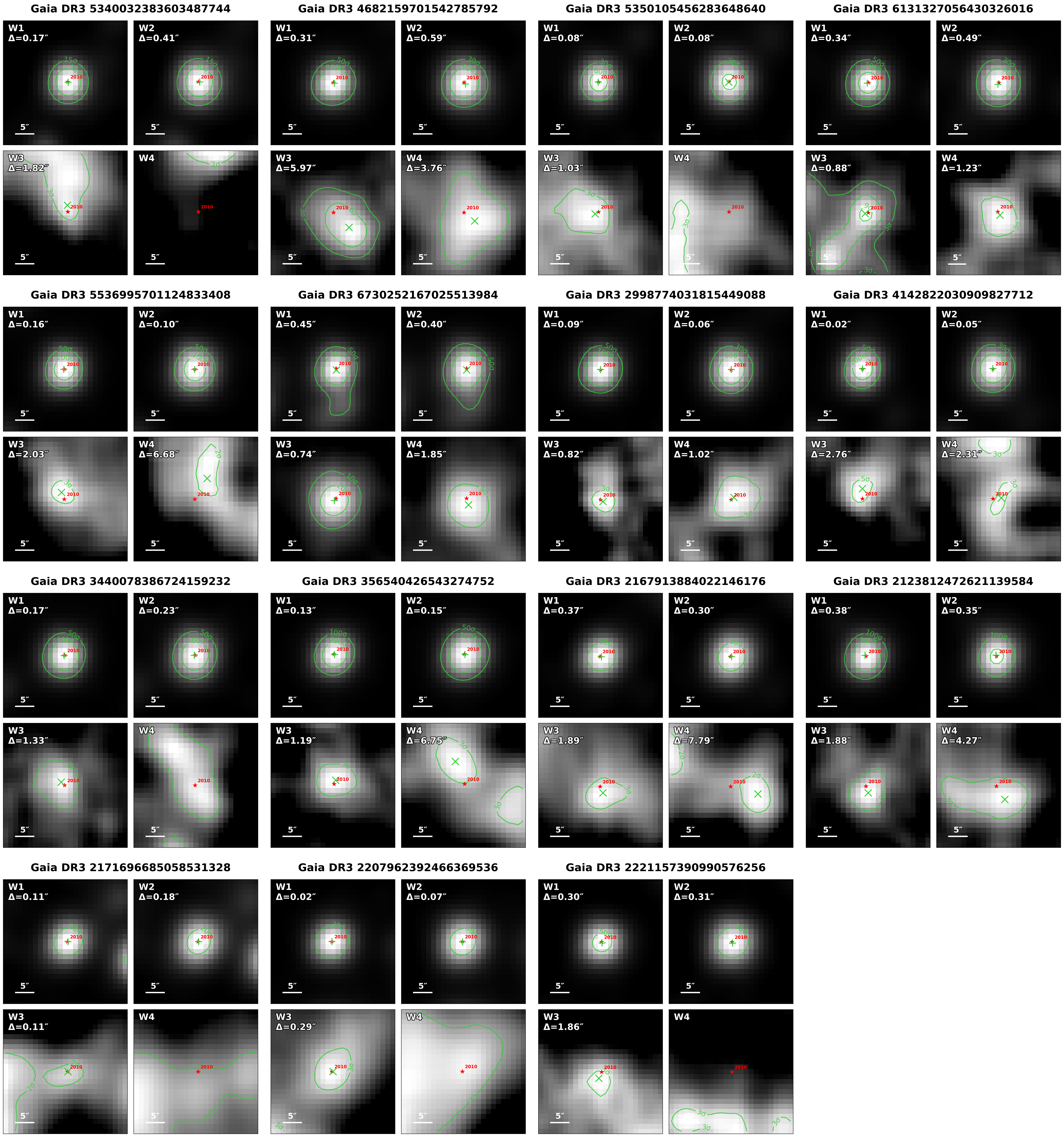}
\caption{ALLWISE infrared images of the 15 control-sample stars for candidate star A. For each star, the four WISE bands are shown, each panel spanning a field of view of $30\arcsec\times30\arcsec$. Green contours show surface brightness at multiple $\sigma$ levels above the background, where $\sigma$ is the per-pixel background noise (RMS), and the green crosses ($\times$) and plus signs (+) mark the MIR centroids. The red star indicates the propagated \emph{Gaia} position at the ALLWISE epoch. The angular offset, $\Delta$, between the propagated \emph{Gaia} position and the measured centroid is labelled in each panel. Note that six control stars lacked significant $W4$ emission and were consequently excluded from the $W4$ centroid fitting.}
\label{fig:control_sample_A}
\end{figure*}

As described in Section~\ref{sec:section_3_5}, we performed the centroid analysis on the full set of 155 control stars, grouped into nine control samples corresponding to the nine candidates. To illustrate the procedure, Figure~\ref{fig:control_sample_A} shows the centroiding results for the 15 control stars of candidate~A as an example. For each star and WISE band, we mark the \emph{Gaia} position propagated to the corresponding WISE observation epoch and the measured centroid, from which the positional offset and associated uncertainty are derived. During the positional analysis of the control stars, the \emph{Gaia} astrometric data are used to propagate the stellar positions to the WISE observation epochs, accounting for proper motion and parallax.

\begin{figure*}
\centering
\includegraphics[width=\textwidth]{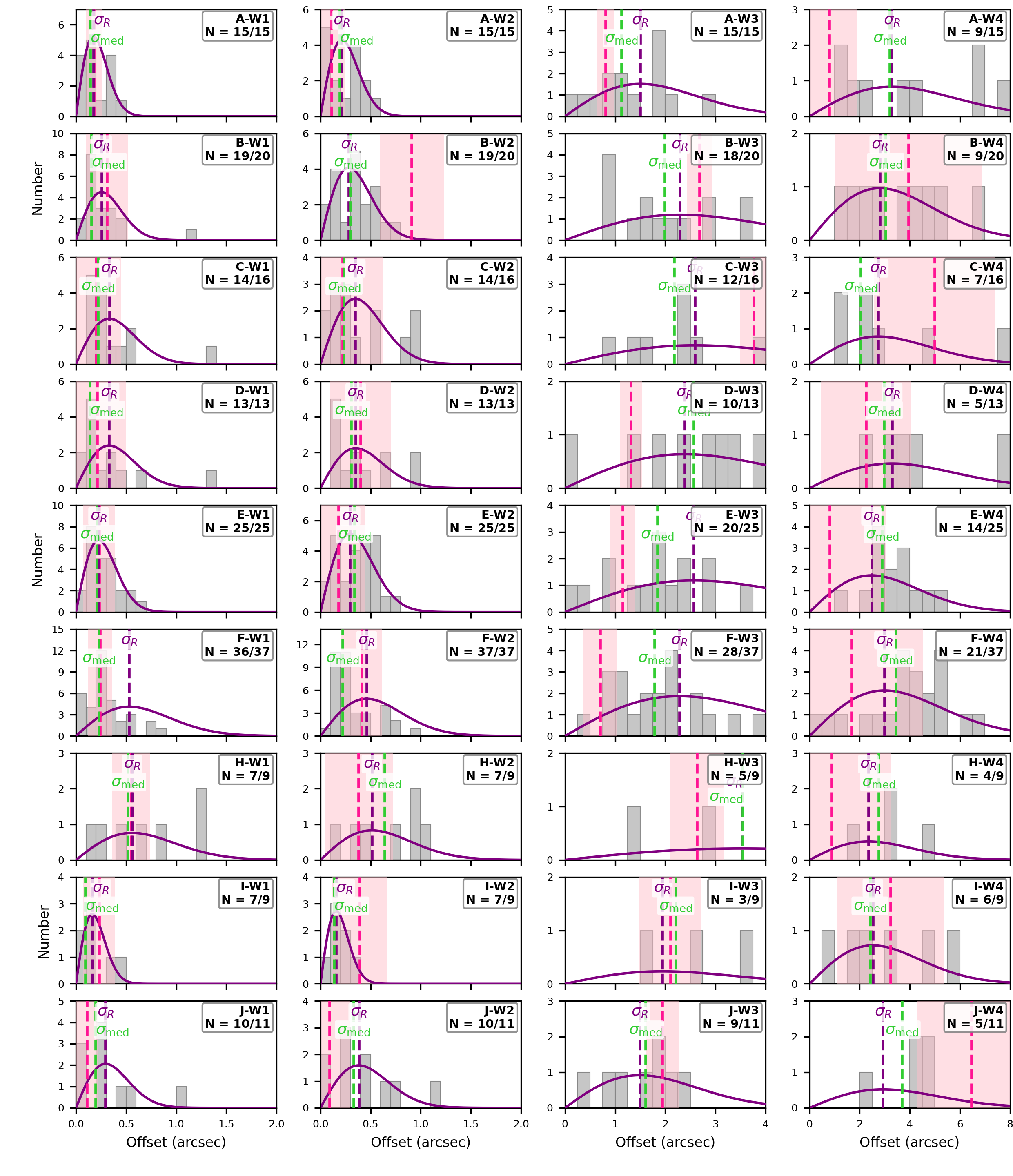}
\caption{Distributions of ALLWISE centroid-to-stellar-position offsets for the control-star samples of the candidates A, B, C, D, E, F, H, I and J, with one candidate per row and the four WISE bands across the columns. In each panel the grey histogram shows the number of control stars per offset bin, with bin widths of 0.1~$\arcsec$ for $W1$ and $W2$, 0.25~$\arcsec$ for $W3$, and 0.5~$\arcsec$ for $W4$. The solid purple curves display the maximum-likelihood Rayleigh distribution fitted to that control sample; the purple dashed lines mark its Rayleigh scale parameter $\sigma_R$, and the limegreen dashed lines show the median-based robust scale $\sigma_{\mathrm{med}}$. The light-red shaded band gives the measured offset of the corresponding candidate star, its half-width the $1\sigma$ centroid uncertainty, and the red dashed line marks the central value. The number of control stars with successful centroid fits is given in each panel as $N=n/n_{\mathrm{tot}}$.}
\label{fig:control_offset}
\end{figure*}

To further address the centroiding analysis presented in Section~\ref{sec:section_3_5}, we provide the complete set of results for all targets in Fig.~\ref{fig:control_offset}. This figure displays the full grid of centroid-offset distributions for the control samples of nine candidates across the four WISE bands. In each panel, a direct visual comparison can be made between the distribution of the background control stars and the specific offset of the candidate. This comprehensive visualisation directly reflects the statistical metrics compiled in Table~\ref{tab:rayleigh}; in particular, candidate~B's $W2$-band offset stands out from its control population.



\bsp	
\label{lastpage}
\end{document}